\documentclass{article}
\usepackage{braket}
\usepackage[
bookmarks=true,bookmarksnumbered=true,
colorlinks = true,anchorcolor= blue, urlcolor=magenta ,
linkcolor = darkblue, citecolor = darkred
]{hyperref}
\usepackage{amsfonts,amsthm,amsmath,amssymb,graphicx}
\usepackage{ulem}
\usepackage{braket}
\usepackage{graphicx} 
\graphicspath{{fig/}}
\usepackage{amsmath}
\usepackage{mathtools}
\usepackage{bm}
\usepackage{amssymb}
\usepackage[top=79pt,bottom=79pt,left=73pt,right=73pt]{geometry}
\usepackage{cases}
\usepackage{float}
\usepackage{xcolor}
\usepackage{braket}
\numberwithin{equation}{section}

\usepackage{color}
    \definecolor{darkgreen}{rgb}{0,0.5,0}
     \definecolor{darkblue}{rgb}{0,0,0.6}
    \definecolor{purple}{rgb}{0.4,0.2,0.7}
    \definecolor{darkred}{rgb}{0.7,0,0}
    \definecolor{airforceblue}{rgb}{0.36, 0.54, 0.66}
    \definecolor{cyan}{rgb}{0.0, 1.0, 1.0}
  	\definecolor{cyan(process)}{rgb}{0.0, 0.72, 0.92}

\newcommand{\chiL}{\chi_{\text{L}}}
\newcommand{\chiR}{\chi_{\text{R}}}
\newcommand{\nuL}{\nu_{\text{L}}}
\newcommand{\nuR}{\nu_{\text{R}}}

\newcommand{\chiLk}{\chi_{\text{L},k}}
\newcommand{\chiRk}{\chi_{\text{R},k}}
\newcommand{\chiLz}{\chi_{\text{L},0}}
\newcommand{\chiRz}{\chi_{\text{R},0}}
\newcommand{\chiLkhalf}{\chi_{\text{L},k-\frac{1}{2}}}
\newcommand{\chiRkhalf}{\chi_{\text{R},k-\frac{1}{2}}}

\newcommand{\aff}[1]{${}^{#1}$}

\begin{document}

\begin{center}
{\Large\bf
Dissipative Majorana Chain---Anomalies and Non-Invertible Symmetries
}\\
\bigskip\bigskip
\bigskip\bigskip
{\large  Hung-Hsuan Teh,\footnote{\tt teh@issp.u-tokyo.ac.jp}}\aff{a,b}
{\large Tokiro Numasawa,\footnote{\tt numasawa@issp.u-tokyo.ac.jp}}\aff{a} \\ 
\bigskip\bigskip
\aff{a}: {\small
\it The Institute for Solid State Physics, The University of Tokyo, Kashiwa 277-8581, Japan
}\\
\aff{b}: {\small
\it Graduate School of Informatics, Nagoya University, Nagoya, 464-0814, Japan
}\\

\end{center}

\begin{abstract}
We study symmetries and anomalies of dissipative Majorana chains described by Lindblad dynamics. 
Using the operator-state mapping, we represent the Lindbladian as a non-Hermitian operator on a doubled Hilbert space. In this formulation, the ket and bra contours admit strong fermion parities and strong translations, while their diagonal combinations define weak symmetries. For even and odd chain lengths, we diagnose anomalies by comparing symmetry algebras with and without weak fermion-parity defects. These algebras retain the lattice precursor of the modulo-$8$ anomaly of the closed Majorana chain and exhibit projective relations between strong fermion parity and weak translation, as well as between weak fermion parity and strong translation, which are intrinsic to open quantum systems.
We establish the IR correspondence between the lattice symmetry algebras and those of the doubled Majorana CFT in four spin sectors, identifying chiral fermion parities emanated from the lattice translations.
Finally, we develop the lattice bosonization of the dissipative chain. For even length with collective energy dephasing, combining the defect and no-defect sectors and imposing GSO projections gives rise to an Ising Lindbladian with non-invertible Kramers--Wannier symmetries descending from the anomalous strong translation and fermion parity of the Majorana chain. For odd length, constructing mutually commuting ket- and bra-sector spin operators instead yields an Ising Lindbladian with a Kramers--Wannier duality defect. Our results provide a controlled framework for extending lattice anomalies and non-invertible symmetries to open quantum systems.
\end{abstract}

\newpage

\tableofcontents

\newpage

\bigskip
\bigskip


\section{Introduction}
Symmetry is a central concept in understanding physical systems \cite{noether1918}, because it provides nontrivial constraints even before one solves the dynamics of the system.
One of the most important associated notions is that of a quantum anomaly.
A quantum anomaly can force the ground state to be nontrivial \cite{tHooft:1979rat,Csaki:1997aw}, for example, by requiring ground-state degeneracy.
Originally, quantum anomalies were formulated in relativistic continuum quantum field theory \cite{Adler:1969gk, Bell:1969ts}.

There is a closely related perspective in lattice many-body systems.
The Lieb--Schultz--Mattis (LSM) theorem and its generalizations show that certain microscopic symmetry realizations are incompatible with a unique symmetric gapped ground state
\cite{Lieb:1961fr,2000oshikawa,2000oshikawa_luttinger,Hastings:2003zx}. 
In modern language, such LSM constraints can be interpreted as lattice manifestations of quantum anomalies. One way to diagnose them is to introduce symmetry defects or twisted boundary conditions. 
In the twisted Hilbert space, the symmetry operators can obey a projective algebra that cannot be removed by local counterterms or local phase redefinitions. 
Recent work has clarified how these lattice anomalies are matched by the 't Hooft anomalies of low-energy continuum field
theories
\cite{Cheng:2022sgb,Seiberg:2024gek,Seifnashri:2023dpa,Aksoy:2023hve,Kapustin:2024rrm,Tu:2025bqf,Pace:2026brj}.

A particularly useful example is the translation-invariant Majorana chain, first proposed by Kitaev \cite{Kitaev:2000nmw} as a model of topological superconductors. 
In the corresponding closed Hamiltonian system, the one-site lattice translation and fermion parity are realized projectively on the Hilbert space. 
At the critical point, the lattice model flows to the $c=1/2$ Majorana conformal field theory. On the low-energy states, the microscopic lattice translation is identified as
\begin{align}
T_{\rm lattice} =
(-1)^{F_\text{L}}e^{2\pi iP/L},
\end{align}
where \(P\) is the continuum momentum and \((-1)^{F_\text{L}}\) is the left-moving chiral fermion parity. Thus the chiral fermion parity of the continuum theory is not an accidental emergent symmetry, but an emanant symmetry originating from microscopic lattice translation.
The projective lattice symmetry algebra therefore provides a lattice precursor of the modulo-$8$ anomaly of the Majorana CFT
\cite{10.21468/SciPostPhys.16.3.064,Seiberg:2024gek}.

This also provides a lattice analogue of the modern interpretation of the Adler--Bell--Jackiw anomaly. 
In gauge theories, an axial symmetry with an ABJ anomaly is no longer an ordinary invertible symmetry after the vector $U(1)$ symmetry is gauged; instead, it can survive as a
non-invertible topological symmetry defect
\cite{Choi:2022jqy,Cordova:2022ieu}. 
Similarly, in the Majorana chain, gauging fermion parity maps the fermionic theory to the bosonic Ising
model, and the Majorana translation symmetry is converted into the
non-invertible Kramers--Wannier duality symmetry. 
In this sense, the Majorana/Ising correspondence gives a $1+1$-dimensional lattice realization of the principle that gauging a symmetry can transform an anomalous invertible symmetry into a non-invertible one.

The preceding discussion concerns closed quantum systems. Many experimentally relevant quantum systems, however, are intrinsically open: their dynamics is described not by unitary Hamiltonian evolution, but by quantum channels or, in the Markovian limit, by Lindblad evolution \cite{lindblad1976generators,gorini1976completely}. This setting is not merely a perturbation to closed systems. Dissipation can destroy quantum coherence, but it can also be used constructively to stabilize nonequilibrium steady states and engineer many-body phases \cite{diehl2008quantum,verstraete2009quantum}. More recently, quantum channels and Lindbladian dynamics have been shown to generate intrinsically mixed-state phases and phase transitions without a closed-system counterpart \cite{sang2024mixed,lessa2025strong,ma2025topological,bao2026mixed,Hsin:2023jqa}. One crucial distinction of open quantum systems is that the notion of symmetry itself becomes richer—a symmetry may act strongly or only weakly \cite{Buca:2012zz,Albert:2014yej}. Thus the Hilbert-space and anomaly-based constraints familiar from closed systems must be reformulated in Liouville space.

The goal of this work is to extend this lattice--continuum anomaly matching to open quantum systems. This raises a basic question: what is the appropriate notion of an anomaly in an open system? 
Unlike closed systems, open quantum systems are not described by Hermitian Hamiltonians acting on state vectors. Their time evolution is generated by Liouvillian superoperators acting on density matrices. 
Moreover, the open-system formulation must incorporate complete positivity, trace preservation, the distinction between strong and weak symmetries, and symmetries intrinsic to the doubled Hilbert space such as modular conjugation
\cite{PRXQuantum.4.030328}. 
While several aspects of open-system anomalies have recently been explored, including LSM-type constraints and anomalous transformation laws of dissipative form factors in Lindbladian systems \cite{Kawabata:2023dlv}, spectral-flow-like diagnostics in open spin chains \cite{zang2024detecting}, and superoperator-based formulations of anomalies for strong and weak symmetries \cite{wang2025anomaly}.
Open system dynamics are also studied in conformal field theories \cite{DelCampo:2019afl,Meng:2020czp,Ishii:2025qpy,Tang:2026axr}
However two issues remain incomplete: a systematic lattice-to-continuum understanding of open-system anomalies and their realizations in fermionic open systems.

We adopt a lattice-to-continuum strategy. 
A finite-dimensional lattice Lindbladian provides a controlled setting in which symmetry operators, symmetry defects, twisted boundary conditions, and allowed local phase redefinitions can be defined explicitly. 
After identifying an anomalous symmetry algebra at finite size, one can take the continuum limit and ask how the same obstruction is reproduced in the corresponding open quantum field theory. 
This strategy is especially natural for the Majorana chain, because the relation between the lattice anomaly and the continuum Majorana anomaly is already well understood in the closed Hamiltonian setting.

In this paper, we study an open version of the translation-invariant
Majorana chain described by Lindblad dynamics. Using the operator-state mapping, we represent the Lindbladian as a non-Hermitian operator acting on a doubled Hilbert space. The ket and bra copies admit strong fermion-parity and translation symmetries, while their diagonal combinations define weak symmetries. 
We analyze the symmetry algebras both in the untwisted theory and in the presence of fermion-parity defects. We identify projective phases that cannot be removed by local redefinitions of the symmetry operators. 
In this work, we refer to such
an obstruction in the symmetry algebra of a twisted Lindbladian as an open-system anomaly.

Our main result is that Lindblad dynamics produces anomaly structures that are not simple copies of the closed-system anomaly. 
In particular, strong and weak symmetries play different roles. While weak diagonal symmetries can sometimes cancel the anomalies of the ket and bra copies, projective algebras involving strong symmetries can remain nontrivial. 
For example, in the doubled Majorana chain, the weak
fermion parity can have a projective algebra with strong translation symmetries. This gives an open-system version of the LSM anomaly, which is intrinsic to the Liouville-space formulation.

We then take the continuum limit and compare the lattice symmetry algebras with those of the doubled Majorana CFT.
We 
analyze the action of fermion parities, chiral fermion parities, translations, and modular conjugation. 
The zero-mode structure in the RR, NSR, and RNS sectors reproduces the projective algebras found on the lattice. 
In this sense, the dissipative Majorana CFT provides the continuum counterpart of the open-chain anomaly.

Finally, we discuss the Jordan--Wigner transformation and bosonization of the dissipative Majorana chain. In the closed Hamiltonian case, summing over spin structures maps the Majorana chain to the Ising model, and the Majorana translation becomes a non-invertible lattice translation symmetry. We explain how this structure is modified in the doubled Hilbert space and discuss its relation to bosonic open Ising systems with non-invertible symmetries.

Rather than proposing a complete classification of anomalies in arbitrary open quantum systems, our aim is to construct a concrete and controlled example of lattice-to-continuum anomaly matching in Lindblad dynamics. 
The dissipative Majorana chain provides such an example: the closed-system lattice anomaly is lifted to Liouville space, where it becomes a projective algebra of strong and weak symmetries, and it
is matched in the continuum by the doubled Majorana field theory.

This paper is organized as follows. In Sec. \ref{sec:Majorana_chain_and_sym_in_Hamiltonian_systems}, we review the symmetries and anomalies of closed Majorana chains and their relation to the Majorana CFT and the Ising model. Section \ref{sec:lindbladian_Maj_chain_and_symmetries} introduces the Lindbladian Majorana chain and its strong and weak symmetries in the doubled Hilbert space. In Sec. \ref{eq:anomaly_of_Lindbladian_Majorana_chain}, we analyze the symmetry algebras and anomalies for even and odd chain lengths, with and without a weak fermion-parity defect. Section \ref{sec:IR_limit_of_dissipative_Majorana_chain} examines the low-energy limit and matches the lattice symmetries with those of the doubled Majorana CFT in the four spin sectors. In Sec. \ref{sec:bosonization}, we develop the bosonization to achieve open Ising systems, including non-invertible symmetries for even chains and a Kramers--Wannier defect for odd chains. We conclude in Sec. \ref{sec:conclusion} with a summary and outlook.

\section{Majorana Chain and Symmetries in Hamiltonian Systems}\label{sec:Majorana_chain_and_sym_in_Hamiltonian_systems}

In this section, we briefly review the symmetry and anomaly structure of the Majorana chain in a closed Hamiltonian system. We focus on the aspects that provide useful background for the Lindbladian generalization below.
In particular, we first recall how fermion parity and chiral fermion parity are realized in different spin sectors of the Majorana CFT. We then explain how the chiral fermion parity arises microscopically from the one-site translation symmetry of the lattice Majorana chain. The resulting lattice symmetry algebra will serve as the closed-system reference for the anomaly analysis in the following sections.

\subsection{Majorana CFT and Spin Structures}
\label{sec:majorana_cft_spin_structures}

We consider the $c=1/2$ Majorana CFT with left- and right-moving Majorana fields $\chiL$ and $\chiR$. The Hamiltonian is
\begin{equation}
  H_{\mathrm{Maj}}
  = i\int dx\,
  \left(\chiL\partial_x\chiL-\chiR\partial_x\chiR\right).
\end{equation}
On a circle of circumference $2\pi$, the fields have the mode expansions
\begin{align}
  \chiL(t,x)
  &= \frac{1}{\sqrt{4\pi}}
  \sum_{r\in\mathbb Z+\nuL}\chi_{\text{L},r}e^{-ir(t+x)},
  \nonumber\\
  \chiR(t,x)
  &= \frac{1}{\sqrt{4\pi}}
  \sum_{r\in\mathbb Z+\nuR}\chi_{\text{R},r}e^{-ir(t-x)},
\end{align}
where
\begin{equation}
  \chi_{a,r}^{\dagger}=\chi_{a,-r},
  \qquad
  \{\chi_{a,r},\chi_{b,r'}\}
  =\delta_{ab}\delta_{r,-r'},
  \qquad a,b\in\{\text{L},\text{R}\}.
\end{equation}

Periodic, or Ramond (R), boundary conditions correspond to $\nu_a=0$, while antiperiodic, or Neveu--Schwarz (NS), boundary conditions correspond to $\nu_a=1/2$. The first and second letters in each sector label refer to the left- and right-moving fields, respectively. Thus RR contains two zero modes, NSNS contains none, and NSR and RNS each contain one.

The total and chiral fermion-parity transformations act as
\begin{align}
  (-1)^F &: (\chiL,\chiR)\rightarrow(-\chiL,-\chiR),\\
  (-1)^{F_\text{L}} &: (\chiL,\chiR)\rightarrow(-\chiL,\chiR),\\
  (-1)^{F_\text{R}} &: (\chiL,\chiR)\rightarrow(\chiL,-\chiR).
\end{align}
Their implementation as operators depends on the spin sector. We work within a single irreducible representation of the fermion algebra in each sector and normalize each implemented parity operator to square to one.

Define the nonzero-mode occupation numbers by
\begin{equation}
  N_a
  =\sum_{\substack{r\in\mathbb Z+\nu_a\\r>0}}
  \chi_{a,-r}\chi_{a,r},
  \qquad a=\text{L},\text{R}.
\end{equation}
Conjugation by $(-1)^{N_a}$ reverses the sign of every nonzero mode of chirality $a$, while leaving all zero modes and all modes of the opposite chirality unchanged.

In the \textbf{NSNS sector}, where $\nuL=\nuR=1/2$ and no zero mode exists, the parity operators can be simply represented as
\begin{equation}
  (-1)^F=(-1)^{N_\text{L}+N_\text{R}},
  \qquad
  (-1)^{F_\text{L}}=(-1)^{N_\text{L}},
  \qquad
  (-1)^{F_\text{R}}=(-1)^{N_\text{R}}.
\end{equation}
They therefore satisfy
\begin{equation}
  (-1)^F(-1)^{F_\text{L}}=(-1)^{F_\text{L}}(-1)^F.
  \label{eq:nsns_linear_algebra}
\end{equation}

In the \textbf{RR sector}, two zero modes exist ($\nuL=\nuR=0$) and they satisfy
\begin{equation}
  \chiLz^{\,2}=\chiRz^{\,2}=\frac{1}{2},
  \qquad
  \{\chiLz,\chiRz\}=0.
  \label{eq:rr_zero_mode_algebra}
\end{equation}
The parity operators can be represented on the full Hilbert space as
\begin{align}
  (-1)^F
  &= -2i\,\chiLz\chiRz(-1)^{N_\text{L}+N_\text{R}},\\
  (-1)^{F_L}
  &= \sqrt2\,\chiRz(-1)^{N_\text{R}},\\
  (-1)^{F_\text{R}}
  &= \sqrt2\,\chiLz(-1)^{N_\text{L}}.
\end{align}
These operators square to one and obey
\begin{equation}
  (-1)^F(-1)^{F_\text{L}}=-(-1)^{F_\text{L}}(-1)^F.
  \label{eq:rr_projective_algebra}
\end{equation}
This anticommutation does not depend on the particular choice of implementers: within a fixed irreducible representation of the fermion algebra, any two operators implementing the same transformation differ only by an overall phase. The symmetry is therefore represented projectively in RR.

These sectors are related by symmetry twists. Taking RR the reference sector, a spatial twist by a symmetry $g$ imposes
\begin{align}
    \chi_a(t,x+2\pi) = g \chi_a(t,x) g^{-1},\qquad
    a=\text{L},\text{R}.
\end{align}
For total fermion parity, $g = (-1)^F$, this gives
\begin{align}
    \chiL(t,x+2\pi) = -\chiL(t,x),\qquad
    \chiR(t,x+2\pi) = -\chiR(t,x),
\end{align}
so the twist changes RR into NSNS. Such twists probe whether the symmetry can be consistently coupled to a background gauge field; an obstruction to doing so is called an anomaly.
The relevant observation concerns the symmetry algebra: total and left-chiral fermion parity commute as transformations of the fields, but their implementing operators commute in NSNS and necessarily anticommute in RR. The latter minus sign cannot be removed by rephasing the operators. This projective realization is a known manifestation of the chiral fermion-parity anomaly\cite{delmastro2021global}.

The \textbf{mixed NSR and RNS sectors} each contain one Majorana zero mode.
Consider NSR, whose zero mode satisfies
\begin{equation}
  \chiRz^{\,2}=\frac{1}{2}.
\end{equation}
On the ground-state subspace, the zero-mode algebra has two 1D
irreducible representations,
\begin{equation}
  \sqrt2\,\chiRz=+1
  \qquad\text{or}\qquad
  \sqrt2\,\chiRz=-1.
\end{equation}
Fermion parity must preserve the ground-state subspace and reverse the zero
mode,
\begin{equation}
  (-1)^F\chiRz \left[ (-1)^F \right]^{-1}=-\chiRz.
\end{equation}
However, conjugation cannot change the sign of a scalar operator within either 1D representation. Thus fermion parity exchanges the two representations and cannot be implemented within either separately. The same obstruction applies to right chiral parity in NSR, since it also reverses $\chiRz$. Similarly, in RNS, total fermion parity and left chiral parity cannot be implemented within a single irreducible representation. Since the mixed sectors can be constructed by introducing a chiral fermion parity defect to the RR sector, this obstruction also signals an anomaly. In fact, when considering general cases with multiple of Majorana fermions, one can show that the obstruction corresponds to a modulo-$2$ anomaly\cite{delmastro2021global, 10.21468/SciPostPhys.16.3.064}.

The mixed sectors also provide a momentum diagnostic of the anomaly. Take NSR sector as a concrete example. Let $\vert\Omega\rangle_\text{NSR}$ denote ground states annihilated by all positive-frequency modes with $r>0$. Its conformal weights are $(h_\text{L},h_\text{R})=(0,1/16)$. On a circle of circumference $2\pi$, a state's momentum $P$ equals its left conformal weight minus its right conformal weight, so
\begin{align}
    P\vert\Omega\rangle_\text{NSR} = -\frac{1}{16} \vert\Omega\rangle_\text{NSR}.
\end{align}
Acting with $\chi_{\text{L},-r}$ increases the momentum by $r$, whereas acting with $\chi_{\text{R},-r}$ decreases it by $r$. In NSR, the allowed positive indices are half-integers on the left and integers on the right. Consequently, all momentum shifts are multiples of $1/2$:
\begin{align}
    P_\text{NSR} \in -\frac{1}{16} + \frac{\mathbb{Z}}{2}.
    \label{eq:P_NSR}
\end{align}
It is known that such fractional momentum offsets in sectors twisted by chiral fermion parity diagnose its $\mathbb{Z}_8$ anomaly. More generally, if one considers $\nu$ copies of Majorana fermions, the momentum selection rule takes the form
\begin{align}
    P \in -\frac{\nu}{16} + \frac{\mathbb{Z}}{2},
\end{align}
and so the offset distinguishes $\nu$ modulo eight\cite{10.21468/SciPostPhys.16.3.064}.

\subsection{Symmetry Anomaly of the Lattice Majorana Chain}
\label{subsec:lattice_majorana_symmetry_anomaly}

We next consider the translation-invariant Majorana chain. The Majorana operators satisfy
\begin{equation}
  \chi_l^\dagger=\chi_l,
  \qquad
  \{\chi_l,\chi_{l'}\}=2\delta_{l,l'},
\end{equation}
and the Hamiltonian is
\begin{equation}
  H=\frac{i}{2}\sum_{l=1}^{L}\chi_{l+1}\chi_l.
\end{equation}
The boundary condition is specified by $\chi_{L+1}=\chi_1$ for the periodic chain and $\chi_{L+1}=-\chi_1$ for the antiperiodic chain.

We now examine the lattice symmetry algebra and its relation to the continuum anomalies discussed above. We begin with an even number of Majorana sites, $L=2N$, without a fermion-parity defect. This corresponds to the RR spin structure in the continuum. As we will see the similar result and detailed derivation in the case of the open quantum systems, the fermion-parity and translation operators satisfy\cite{PhysRevLett.117.166802,10.21468/SciPostPhys.16.3.064}
\begin{equation}
  (-1)^F T_{\text{RR}}
  =
  -T_{\text{RR}}(-1)^F.
\end{equation}
This is the lattice counterpart of the projective relation between $(-1)^F$ and $(-1)^{F_L}$ in Eq.~\eqref{eq:rr_projective_algebra}.

We can probe the same anomaly structure by inserting a fermion-parity defect. On the lattice, this changes the boundary condition of the Majorana fermions from periodic to antiperiodic. The corresponding continuum theory is then in the
NSNS sector. In this sector the symmetry algebra becomes
\begin{equation}
  (-1)^F T_{\mathrm{NSNS}}
  =
  T_{\mathrm{NSNS}}(-1)^F.
\end{equation}
Thus the symmetry realization differs between the no-defect RR sector and the defect NSNS sector. This change of the symmetry algebra under a fermion-parity twist is the closed-system prototype of the defect-based anomaly diagnosis that we will later apply to the Lindbladian Majorana chain.

The situation is different when the number of lattice Majoranas is odd, $L=2N+1$. In this case the product of all Majorana operators,
\begin{equation}
  \mathcal{C}=\chi_1\chi_2\cdots\chi_{2N+1},
  \label{eq:odd_majorana_central_product}
\end{equation}
commutes with every operator in the Clifford algebra. Within an irreducible representation, $\mathcal{C}$ is therefore proportional to the identity and acts as a fixed nonzero scalar. Fermion parity would transform every Majorana as $\chi_\ell\longrightarrow-\chi_\ell$, and hence
\begin{equation}
  \mathcal{C}\longrightarrow-\mathcal{C}.
\end{equation}
It therefore maps one irreducible representation to another rather than acting within a single irreducible Hilbert space. This is the lattice version of the single-zero-mode obstruction discussed in
Sec.~\ref{sec:majorana_cft_spin_structures}.

The low-energy theory of the odd chain corresponds to the NSR sector. Although fermion parity itself cannot be implemented within a single irreducible Hilbert space, the translation operator is well defined. Its $L$-th power $T_\text{NSR}^L$ initially carries a phase that depends on the chain length. Following the local-counterterm prescription of Ref. \cite{10.21468/SciPostPhys.16.3.064}, we allow phase redefinitions $T_\text{NSR} \rightarrow e^{i(\alpha L + \beta)} T_\text{NSR}$, with $\alpha$, $\beta$ independent of $L$. Choosing these parameters to remove the length dependence yields 
\begin{align}
    T_\text{NSR}^L = e^{i 2\pi n / 16},\qquad
    n \in 2\mathbb{Z}+1.
    \label{eq:T_NSR_to_L_closed_sys}
\end{align}
Such redefinitions can change $n$, but not its oddness. This constitutes a modulo-$2$ lattice obstruction.

The connection to the continuum is provided by a crucial lattice-continuum correspondence,
\begin{align}
    T \sim (-1)^{F_\text{L}} e^{i 2\pi P/L}.
    \label{eq:dictionary_closed_sys}
\end{align}
With the overall phase fixed by the low-energy normalization\cite{10.21468/SciPostPhys.16.3.064}, this correspondence gives
\begin{align}
    T_\text{NSR}^L = e^{-i 2\pi / 16},
\end{align}
corresponding to $n=-1$ and consistent with the NSR momentum spectrum in Eq. (\ref{eq:P_NSR}). Notice that while the lattice algebra determines only the oddness of $n$, the low-energy reveals the finer continuum $\mathbb{Z}_8$ anomaly.
For readers who are interested in details for Eq. (\ref{eq:dictionary_closed_sys}), please refer to Ref. \cite{10.21468/SciPostPhys.16.3.064} or Sec. \ref{sec:IR_limit_of_dissipative_Majorana_chain}, where we develop a generalized lattice-continuum correspondence for open quantum systems.

The closed-system results relevant to the remainder of the paper can therefore be summarized as follows. For even $L$, the anomaly is visible through the different projective symmetry algebras in the RR and NSNS sectors, related by
the insertion of a fermion-parity defect. For odd $L$, fermion parity cannot be implemented within a single irreducible Hilbert space, while the anomaly is encoded in the fractional phase of $T^L$.

In the open system, these structures are modified because fermion parity and translation admit separate strong actions on the ket and bra sectors as well as diagonal weak actions. In particular, we will introduce a \emph{weak
fermion-parity defect}, which twists both copies of the doubled Hilbert space. For even $L$, the no-defect and defect theories correspond to the RR and NSNS spin structures, respectively. For odd $L$, the no-defect theory gives the NSR
structure, while the weakly twisted doubled theory gives the RNS structure discussed later. The aim of the following sections is to determine how the closed-system anomaly relations above are modified by the distinction between strong and weak symmetries.

\subsection{From Majorana translation to non-invertible Ising symmetry}
\label{subsec:majorana_noninvertible_ising_symmetry}

An anomalous invertible symmetry has been shown to give rise to a non-invertible symmetry after bosonization. The Majorana chain provides a concrete example. For an even chain with $L=2N$, bosonization utilizes the Jordan--Wigner transformation, combines the RR and NSNS Hilbert spaces, $\mathcal{H}_\text{NSNS} \oplus \mathcal{H}_\text{RR}$, and applies a projection retaining the even-parity NSNS states and odd-parity RR states. This gives the critical Ising Hamiltonian,
\begin{equation}
  H_{\mathrm{Ising}}
  =-\frac12\sum_{j=1}^{N}\left(Z_j+X_jX_{j+1}\right),
  \qquad X_{N+1}=X_1.
\end{equation}
Now, we discuss how symmetries of the Majorana chains get transformed under the bosonization. Before the projection, the one-site Majorana translations are invertible. The translation in NSNS preserves the retained parity subspace. However, translation in RR anticommutes with fermion parity and therefore takes the retained subspace outside the projected Hilbert space. One thus naturally constructs an operator that agrees with translation in NSNS and vanishes in RR, i.e.
\begin{align}
    \widetilde{\mathsf{D}} =
    \begin{pmatrix}
        T_\text{NSNS} & 0\\
        0 & 0
    \end{pmatrix},
\end{align}
which is non-invertible because of the kernel. It preserves the projected Hilbert space and, after the projection, becomes the non-invertible Kramers--Wannier symmetry $\mathsf{D}$, satisfying
\begin{align}
    \mathsf D Z_j=X_jX_{j+1}\mathsf D,
    \qquad
    \mathsf D X_jX_{j+1}=Z_{j+1}\mathsf D.
\end{align}
Consequently,
\begin{equation}
  \mathsf D H_{\mathrm{Ising}}=H_{\mathrm{Ising}}\mathsf D.
\end{equation}

We develop the corresponding bosonization and symmetry construction for open quantum systems in Sec. \ref{sec:bosonization}. Although formulated in the doubled Hilbert space, those derivations also explain the closed-system mechanism summarized here. There, we demonstrate that for an even number of Majorana sites, the anomalous relations between strong fermion parity and strong translation in the RR sector lead, after bosonization, to non-invertible Kramers--Wannier symmetries of the Ising Lindbladian. For an odd number of Majorana sites, however, the analogous construction does not produce such symmetries, despite the anomalous parity-translation relations in the RNS sector. The bosonized model instead, similar to the closed Majorana chain, contains a Kramers--Wannier duality defect. These contrasting outcomes illustrate how the doubled-space symmetry structure modifies the relation between anomalies and non-invertible symmetries.

\section{Lindbladian Majorana Chains and Symmetries}\label{sec:lindbladian_Maj_chain_and_symmetries}
We consider a Majorana chain coupled to an environment, whose Markovian dynamics is governed by the Lindblad equation,
\begin{align}
    &\frac{d\rho}{dt}=-i[H,\rho] + \sum_{m}\left(
    L_{m} \rho L_{m}^{\dagger}  -  \frac{1}{2}\{L_{m}^{\dagger}L_{m},\rho\}  \right),\label{eq:Lindblad_eq}
\end{align}
where $\rho$ represents the density matrix, $H$ is the Hamiltonian, and $\{L_{m}\}$ is a set of jump operators describing the coupling between the Majorana chain and the environment. Both $H$ and $\{L_{m}\}$ are composed of Majorana fermions $\chi_{l}$, with site index $l$ running from $1$ to $L$, satisfying $\chi_{l}^{\dagger} = \chi_{l}$ and the anticommutation relation,
\begin{align}
    \{\chi_{l},\chi_{l'}\} = 2\delta_{ll'}.
\end{align}
Although most of our discussion depends only symmetries, it is useful to keep concrete representatives in mind. We will consider the following Majorana chain Hamiltonian:
\begin{align}
    H_0 &= \frac{i}{2}\sum_{l=1}^{L-1} \chi_{l+1}\chi_l + \frac{i}{2} \chi_1 \chi_L,\label{eq:H0_Majorana_chain}
\end{align}
where the subscript $0$ denotes no-defect.

As a concrete example of dissipation, we mostly consider a single collective energy-dephasing jump operator,
\begin{align}
    L_0 = \sqrt{\gamma} \left( i\sum_{l=1}^{L-1} \chi_{l+1}\chi_l + i \chi_1 \chi_L \right),\label{eq:L0_collective_density_dephasing_jump_op}
\end{align}
where $\gamma$ controls the dissipative strength. Since there is only one jump operator in this example, we suppress the jump-operator index $m$. Physically, such a collective dephasing channel may arise from coupling the Majorana chain to a fluctuating cavity mode that modulates the energy density. Other types of jump operators will be introduced in Sec. \ref{sec:jump_operator_examples}.

\subsection{Doubled Hilbert Space and Lindbladian}
To investigate the symmetries of the Lindblad equation, we employ the doubled Hilbert space formalism and introduce two types of Majorana fermions, $\chi_{l}^{+}$ and $\chi_{l}^{-}$, acting on the ket and bra spaces, respectively. Following Ref. \cite{Kulkarni:2021gtt,Kawabata:2022osw,PRXQuantum.4.030328}, we impose the anticommutation relation between operators in different sectors. That is,
\begin{align}
    \{\chi_{l}^{s},\chi_{l'}^{s'}\} = 2\delta_{ll'}\delta_{ss'}.
\end{align}
The ket- and bra-space Majorana fermions are glued through the maximally entangled state $\vert I\rangle$ via \cite{Kulkarni:2021gtt,Kawabata:2022osw,PRXQuantum.4.030328}
\begin{align}
    \chi_{l}^{+}  \vert I\rangle
    = -i \chi_{l}^{-} \vert I\rangle.
\end{align}
The reference state $\ket{I}$ plays a role of a reflector. Note that $\vert I\rangle$ is the operator-to-state mapping of the identity operator $I$ in the Hilbert space. Thus, for any ket-space operator $A^{+}$, $A^{+}\vert I\rangle = \vert AI\rangle = \vert A\rangle$.

For any linear operator $A$ generated by Majorana fermions, i.e. $A(\{\chi_l\})$, in the Hilbert space, we can construct a ket-space representative as
\begin{align}
    A^+ = A(\{\chi_l^+\}).\label{eq:def_A_plus}
\end{align}
Here, however, the designation ``ket-space'' should not be confused with the set of Majorana generators from which an operator is constructed. An operator whose action is nontrivial only on the ket Majoranas $\chi_l^+$ need not be generated solely by $\left\{\chi_l^+\right\}$. In fact, as we will see in Sec. \ref{sec:fermion_parity_and_topol_defect}, for an odd number of sites the ket-space fermion-parity is expressed entirely in terms of the bra-space Majoranas $\chi_l^-$.

The bra-space correspondence is defined using the modular conjugation $\mathcal{J}$, an antiunitary operator that arises from Hermiticity preservation and exchanges ket and bra representatives (see Sec. \ref{sec:modular_conjugation} for more details),
\begin{align}
    A^- \equiv \mathcal{J} A^+ \mathcal{J}^{-1}, \quad A^+ = \mathcal{J} A^- \mathcal{J}^{-1}.\label{eq:def_A_minus}
\end{align}
As an illustration, the ket and bra representatives of Eqs. (\ref{eq:H0_Majorana_chain}) and (\ref{eq:L0_collective_density_dephasing_jump_op}) are
\begin{align}
    H_0^\pm &= \pm \frac{i}{2}\sum_{l=1}^{L-1} \chi_{l+1}^\pm \chi_l^\pm \pm \frac{i}{2} \chi_1^\pm \chi_L^\pm,\label{eq:H_0_plus_minus}\\
    L_0^\pm &= \sqrt{\gamma} \left( \pm i\sum_{l=1}^{L-1} \chi_{l+1}^\pm \chi_l^\pm \pm i \chi_1^\pm \chi_L^\pm \right).
\end{align}

We now rewrite the Lindblad equation, Eq. (\ref{eq:Lindblad_eq}), in the doubled Hilbert space by mapping it to the ket space, followed by operating on the reference state $\vert I\rangle$. As an illustration, consider the right-multiplication term $i\rho H$ in Eq. (\ref{eq:Lindblad_eq}) with the Majorana chain Hamiltonian $H_0$. Using Eqs. (\ref{eq:def_A_plus})-(\ref{eq:H_0_plus_minus}), we obtain
\begin{align}
    i\rho H_0 \rightarrow (i\rho H_0)^{+} \vert I\rangle &= i \rho^{+}H_0^{+} \vert I\rangle
    = i \rho^{+} \frac{i}{2} \sum_l \chi_{l+1}^+ \chi_l^+ \vert I\rangle
    = \rho^+ \frac{i}{2} \sum_l \chi_{l+1}^+ \chi_l^- \vert I\rangle \notag\\
    &= - \rho^+ \frac{i}{2} \sum_l \chi_l^- \chi_{l+1}^+ \vert I\rangle
    = i \rho^+ \frac{i}{2} \sum_l \chi_l^- \chi_{l+1}^- \vert I\rangle
    = - i \frac{i}{2} \sum_l \chi_{l+1}^- \chi_{l}^- \rho^+ \vert I\rangle
    = i H_0^- \vert\rho\rangle.
\end{align}
For fermion-parity even $H$, and for $L_m$ with definite fermion parity, the overall result takes a compact Schr\"{o}dinger-like form,
\begin{align}
    \frac{d\vert\rho\rangle}{dt} =&
    \begin{cases}
        -i (H^{+}-H^{-})\vert\rho\rangle
        +\sum_{m}\left(  -iL_{m}^{+}L_{m}^{-}  -\frac{1}{2} L_{m}^{+\dagger}L_{m}^{+}  -\frac{1}{2}L_{m}^{-\dagger}L_{m}^{-}
        \right)\vert\rho\rangle, & \text{fermion-parity odd }\{L_{m}\},\\
        -i (H^{+}-H^{-})\vert\rho\rangle
        +\sum_{m}\left(  L_{m}^{+}L_{m}^{-}  -\frac{1}{2} L_{m}^{+\dagger}L_{m}^{+}  -\frac{1}{2}L_{m}^{-\dagger}L_{m}^{-}
        \right)\vert\rho\rangle, & \text{fermion-parity even }\{L_{m}\},
    \end{cases}\label{eq:Lindbladian_in_doubled_Hilbert_space}\\
    \equiv&\mathcal{L} \vert\rho\rangle.
\end{align}
We can therefore explore the symmetries of the Lindbladian $\mathcal{L}$ in the conventional way,
\begin{align}
    U\mathcal{L}U^{-1}=\mathcal{L}.
\end{align}
We say $U$ is a strong symmetry if it acts solely on either the ket or bra space, and a weak symmetry otherwise.

Before discussing different symmetries, we provide an explicit spin representation for the Majorana fermions $\{\chi_{l}^{\pm}\}$, which in practice can be used for verifying all symmetry operations and computing the Lindbladian spectra. In total we have $2L$ Majorana fermions. We choose a Jordan--Wigner transformation in which the Majorana fermions on odd sites to be purely real and those on even sites to be purely imaginary. For even $L=2N$, we take
\begin{align}
    &\chi_{1}^{+}=-X_{1}\\
    &\chi_{2}^{+}=Y_{1}\\
    &\chi_{3}^{+}=-Z_{1}X_{2}\\
    &\chi_{4}^{+}=Z_{1}Y_{2}\\
    &\vdots\\
    &\chi_{2N-1}^{+}=-Z_{1}Z_{2}\cdots Z_{N-1}X_{N}\\
    &\chi_{2N}^{+}=Z_{1}Z_{2}\cdots Z_{N-1}Y_{N}\\
    &\chi_{1}^{-}=(Z_{1}Z_{2}\cdots Z_{N})X_{N+1}\\
    &\chi_{2}^{-}=(Z_{1}Z_{2}\cdots Z_{N})Y_{N+1}\\
    &\chi_{3}^{-}=(Z_{1}Z_{2}\cdots Z_{N})Z_{N+1}X_{N+2}\\
    &\chi_{4}^{-}=(Z_{1}Z_{2}\cdots Z_{N})Z_{N+1}Y_{N+2}\\
    &\vdots\\
    &\chi_{2N-1}^{-}=(Z_{1}Z_{2}\cdots Z_{N})Z_{N+1}\cdots Z_{2N-1}X_{2N}\\
    &\chi_{2N}^{-}=(Z_{1}Z_{2}\cdots Z_{N})Z_{N+1}\cdots Z_{2N-1}Y_{2N}.
\end{align}
Note the representations has dimension $2^{2N}=2^{L}$.\\
For odd $L=2N+1$, we use the same Jordan--Wigner transformation, supplemented by the following for the additional two Majorana fermions:
\begin{align}
    &\chi_{2N+1}^{+}=-(Z_{1}Z_{2}\cdots Z_{N})(Z_{N+1}\cdots Z_{2N})X_{2N+1}\\
    &\chi_{2N+1}^{-}=(Z_{1}Z_{2}\cdots Z_{N})(Z_{N+1}\cdots Z_{2N})Z_{2N+1}.
\end{align}
In this case, the representation has dimension $2^{2N+1}=2^{L}$.

\subsection{Strong and Weak Symmetries}

Symmetry of an open quantum system is naturally formulated as symmetry of the
Lindbladian superoperator. After the operator-state mapping, a unitary symmetry is
represented by an operator acting on the doubled Hilbert space. Let $U^+$ and $U^-$
denote the implementations of a symmetry transformation on the ket and bra Hilbert
spaces, respectively. 

A symmetry acting simultaneously on the ket and bra spaces is called a
\emph{weak symmetry} \cite{Buca:2012zz,Albert:2014yej}. More precisely, let
\begin{align}
U_{\rm w} \propto U^+ U^-
\end{align}
denote the diagonal action on the doubled Hilbert space. The proportionality symbol
allows for an overall phase, which will be fixed later according to the desired group
relations. The weak-symmetry condition is
\begin{align}
U_{\rm w}\mathcal{L}U_{\rm w}^{-1}
=
\mathcal{L}.
\label{eq:weak_symmetry_definition}
\end{align}
In the original density-matrix formulation, this symmetry corresponds to the covariance
of the Lindblad equation under
\begin{align}
\rho \longmapsto U^{\dagger}\rho U ,
\end{align}
where $U$ is the corresponding unitary operator on the original Hilbert space. 
In other words, when we view $\mathcal{L}$ as a superoperator, the weak symmetry implies 
\begin{align}
U^{\dagger}\mathcal{L} (U\cdot U^{\dagger}) U = \mathcal{L} (\cdot ) .
\end{align}
Thus, a weak symmetry is a symmetry of the open-system dynamics as a whole, but it does not necessarily act independently on the ket and bra contours.

By contrast, the symmetry is called a \emph{strong symmetry} if the ket and bra
actions are separately symmetries of the Lindbladian:
\begin{align}
U^+\mathcal{L}(U^+)^{-1}
&=
\mathcal{L},
\notag\\
U^-\mathcal{L}(U^-)^{-1}
&=
\mathcal{L}.
\label{eq:strong_symmetry_definition}
\end{align}
When we view $\mathcal{L}$ as a superoperator, then the strong symmetry implies 
\begin{align}
U^{\dagger}\mathcal{L} (U\cdot )  = \mathcal{L} (\cdot ) .
\end{align}
The strong-symmetry condition is therefore more restrictive than the weak-symmetry
condition. In particular, a strong symmetry always gives rise to a weak symmetry,
whereas a weak symmetry need not be strong \cite{PRXQuantum.4.030328}.
In the Hamiltonian system, what we usually call symmetries correspond to strong symmetries.

\subsection{Modular Conjugation}\label{sec:modular_conjugation}
The modular conjugation, a symmetry intrinsic to open quantum systems, exchanges the two sectors in the double Hilbert space,
\begin{align}
    \mathcal{J}\chi_{l}^{+}\mathcal{J}^{-1}=\chi_{l}^{-},\quad
    \mathcal{J}\chi_{l}^{-}\mathcal{J}^{-1}=\chi_{l}^{+},\quad
    \mathcal{J}z\mathcal{J}^{-1}=z^{*}.
\end{align}
One explicit representation of $\mathcal{J}$ is
\begin{align}
    \mathcal{J}=\left[
    \prod_{l=1}^{N} \frac{1}{\sqrt{2}} \left( \chi_{2l-1}^{+} - \chi_{2l-1}^{-} \right)
    \frac{i}{\sqrt{2}} \left(\chi_{2l}^{+}+\chi_{2l}^{-}\right)
    \right]
    \mathcal{K}
\end{align}
for even $L=2N$, and
\begin{align}
    \mathcal{J}=\left[
    \prod_{l=1}^{N} \frac{1}{\sqrt{2}} \left( \chi_{2l-1}^{+} + \chi_{2l-1}^{-} \right)
    \frac{i}{\sqrt{2}} \left(\chi_{2l}^{+}-\chi_{2l}^{-}\right)
    \right]
    \frac{1}{\sqrt{2}} \left(\chi_{2N+1}^{+}+\chi_{2N+1}^{-}\right)
    \mathcal{K}
\end{align}
for odd $L=2N+1$. Note that $\mathcal{J}^{2}=1$. We also remark that $[\mathcal{L},\mathcal{J}]=0$ in general, irrespective of whether $L_{m}$ is fermion-parity odd or even. This implies a mirror symmetry of the Lindbladian spectrum with respect to the real axis.

\subsection{Fermion Parity and Topological Defect}\label{sec:fermion_parity_and_topol_defect}
Fermion parity changes the signs of the Majorana fermions. There are two strong fermion parities, $(-1)^{F^{+}}$ and $(-1)^{F^{-}}$, satisfying
\begin{align}
    &(-1)^{F^{+}}:\quad \chi_{l}^{+}\rightarrow -\chi_{l}^{+},\quad \chi_{l}^{-}\rightarrow\chi_{l}^{-};\\
    &(-1)^{F^{-}}:\quad \chi_{l}^{+}\rightarrow \chi_{l}^{+},\quad \chi_{l}^{-}\rightarrow -\chi_{l}^{-},
\end{align}
and one weak fermion parity symmetry, $(-1)^{F^\text{w}}$, satisfying
\begin{align}
    (-1)^{F^\text{w}}:\quad \chi_{l}^{\pm}\rightarrow -\chi_{l}^{\pm}.
\end{align}

The distinction between strong and weak symmetries is particularly important for
fermion parity. If every jump operator is even under fermion parity,
\begin{align}
(-1)^F L_m (-1)^F = L_m ,
\end{align}
the fermion parities acting separately on the ket and bra spaces can be strong
symmetries. By contrast, if a jump operator is odd,
\begin{align}
(-1)^F L_m (-1)^F = -L_m ,
\end{align}
the two minus signs cancel in the full dissipator. Consequently, the diagonal fermion
parity acting on both contours remains a weak symmetry, even though the individual
ket- and bra-space fermion parities are generally not strong symmetries.

Explicit representations for the strong symmetries are
\begin{align}
    &(-1)^{F^{+}}=
        \prod_{l=1}^{N}
        \left(
        i\chi^{+}_{2l-1}\chi^{+}_{2l}
        \right),\label{eq:strong_fermion_parity_plus_even}\\
    &(-1)^{F^{-}}=
        \prod_{l=1}^{N}
        \left(
        -i\chi^{-}_{2l-1}\chi^{-}_{2l}
        \right),\label{eq:strong_fermion_parity_minus_even}
\end{align}
when $L=2N$ is even, while
\begin{align}
    &(-1)^{F^{+}}=\left[\prod_{l=1}^{N}
        \left(
        -i\chi^{-}_{2l-1}\chi^{-}_{2l}
        \right)
    \right]
    \chi_{2N+1}^{-},\label{eq:strong_fermion_parity_plus_odd}\\
    &(-1)^{F^{-}}=\left[\prod_{l=1}^{N}
        \left(
        i\chi^{+}_{2l-1}\chi^{+}_{2l}
        \right)
    \right]
    \chi_{2N+1}^{+},\label{eq:strong_fermion_parity_minus_odd} 
\end{align}
when $L=2N+1$ is odd. Unlike in the closed Majorana system, the fermion parities $(-1)^{F^{\pm}}$ for odd $L$ are inner automorphisms, and can be realized in Hilbert space with the aid of Majorana fermions from the opposite sector. A natural extension to the weak symmetry is
\begin{align}
    (-1)^{F^\text{w}}=\prod_{l=1}^{N}
        \left(
        i\chi^{+}_{2l-1}\chi^{+}_{2l}
        \right)
        \left(
        -i\chi^{-}_{2l-1}\chi^{-}_{2l}
        \right)
        = (-1)^{F^{+}} (-1)^{F^{-}}
\end{align}
for even $L=2N$, and
\begin{align}
    (-1)^{F^\text{w}}=\left[\prod_{l=1}^{N}
        \left(
        i\chi^{+}_{2l-1}\chi^{+}_{2l}
        \right)
        \left(
        -i\chi^{-}_{2l-1}\chi^{-}_{2l}
        \right) 
    \right]
    \left( i\chi_{2N+1}^{+}\chi_{2N+1}^{-} \right)
    = -i (-1)^{F^{+}} (-1)^{F^{-}}\label{eq:fermion_parities_projective_rep}
\end{align}
for odd $L=2N+1$. We remark that $\left[(-1)^{F^{\pm}}\right]^{2}=\left[(-1)^{F^\text{w}}\right]^{2}=1$ for both odd and even $L$, and in the odd $L$ case the $\mathbb{Z}_{2}\times\mathbb{Z}_{2}$ type symmetry is realized projectively.

Later to extract the anomaly structure of the Lindbladian Majorana chain, and to construct the corresponding Ising type Lindbladian, we need to introduce twisted sectors. There are two natural ways to achieve this. In the first route, we insert a topological defect directly in the original Hilbert space, replacing $H_0$ and $L_0$ by the twisted counterparts $H_{\mathsf{G}}$ and $L_\mathsf{G}$. We then lift these operators to the doubled Hilbert space to obtain $H_{\mathsf{G}}^\pm$ and $L_{\mathsf{G}}^\pm$ using Eqs. (\ref{eq:def_A_plus}) and (\ref{eq:def_A_minus}). In the second route, we directly impose boundary twists to $H_0^{\pm}$ and $L_0^{\pm}$ in the doubled Hilbert space, obtaining $H_\mathsf{G}^{\pm}$ and $L_\mathsf{G}^{\pm}$. We remark that the twist generated from the first route is equivalent to a ``weak'' twist through the second route---weak in the sense that both copies in the doubled Hilbert space have topological defects. The second route is more flexible, since it also allows one, at least formally, to introduce a strong twist acting on only one copy of the doubled Hilbert space.

In this work, we will focus on the weak fermion-parity twist realized by (following the second route)
\begin{align}
    \chi_{l+L}^{\pm} = (-1)^{F^\text{w}} \chi_{l}^{\pm} (-1)^{F^\text{w}} = -\chi_{l}^{\pm}.
\end{align}
Although the particular forms of the Hamiltonian and the jump operator are not essential, a concrete example of the twisted theory is given by the Majorana chain in Eq. (\ref{eq:H0_Majorana_chain}) together with the jump operator in Eq. (\ref{eq:L0_collective_density_dephasing_jump_op}),
\begin{align}
    H_\mathsf{G}^\pm &= \pm \frac{i}{2}\sum_{l=1}^{L-1} \chi_{l+1}^\pm\chi_l^\pm \mp \frac{i}{2} \chi_1^\pm \chi_L^\pm,\label{eq:H_G}\\
    L_\mathsf{G}^\pm &= \sqrt{\gamma} \left( \pm i \sum_{l=1}^{L-1} \chi_{l+1}^\pm\chi_l^\pm \mp i \chi_1^\pm \chi_L^\pm \right),
\end{align}
where the subscript $\mathsf{G}$ indicates the presence of the defect.

While it is also natural to consider a strong fermion-parity defect using the second route, which would lead to for instance $\chi_{L+1}^{+}=-\chi_{1}^{+}$ in the plus sector and $\chi_{L+1}^{-}=\chi_{1}^{-}$ in the minus sector, the corresponding Lindbladian cannot be realized by a single set of $H$ and $\{L_{m}\}$. Therefore, in this work we will not consider this case.

\subsection{Translation Symmetry}\label{sec:translation_symm}
In closed systems, the translation symmetry $T$ simply transforms $\chi_{l}\rightarrow\chi_{l+1}$, and $\chi_{L+1}$ is identified with $\chi_{1}$. In contrast, in open quantum systems, there are two strong translation symmetries,
\begin{align}
    &T^{+}:\quad \chi_{l}^{+}\rightarrow\chi_{l+1}^{+},\quad \chi_{l}^{-}\rightarrow\chi_{l}^{-};\\
    &T^{-}:\quad \chi_{l}^{+}\rightarrow \chi_{l}^{+},\quad \chi_{l}^{-}\rightarrow\chi_{l+1}^{-},
\end{align}
and one weak translation symmetry $T^{\text{w}}$ satisfying
\begin{align}
    T^{\text{w}}:\quad \chi_{l}^{\pm}\rightarrow\chi_{l+1}^{\pm},
\end{align}
for all $l$ in the untwisted theory.

If the Hamiltonian and each jump
operator are individually invariant under translation, the ket and bra translations can
be strong symmetries. If translation instead permutes the jump operators, the full
Lindbladian is still invariant under the simultaneous translation of the two contours,
while the individual contour translations are generally not symmetries.

When $L=2N$ is even (with no defect), the translation symmetries can be implemented as
\begin{align}
    &T_{\text{RR}}^{+}=\frac{  e^{i\frac{N-1}{4}\pi}  }{  2^{\frac{2N-1}{2}}  }
    \chi_{1}^{+}(1+\chi_{1}^{+}\chi_{2}^{+})(1+\chi_{2}^{+}\chi_{3}^{+})\cdots
    (1+\chi_{2N-1}^{+}\chi_{2N}^{+})(-1)^{F^{-}},\label{eq:TRR_plus}\\
    &T_{\text{RR}}^{-}=\frac{  e^{-i\frac{N-1}{4}\pi}  }{  2^{\frac{2N-1}{2}}  }
    \chi_{1}^{-}(1+\chi_{1}^{-}\chi_{2}^{-})(1+\chi_{2}^{-}\chi_{3}^{-})\cdots
    (1+\chi_{2N-1}^{-}\chi_{2N}^{-})(-1)^{F^{+}},\\
    &T_{\text{RR}}^{\text{w}} = \frac{ i }{  2^{2N-1}  }
    \chi_{1}^{+}\chi_{1}^{-} (1+\chi_{1}^{+}\chi_{2}^{+})(1+\chi_{1}^{-}\chi_{2}^{-}) (1+\chi_{2}^{+}\chi_{3}^{+}) (1+\chi_{2}^{-}\chi_{3}^{-})\cdots
    (1+\chi_{2N-1}^{+}\chi_{2N}^{+})(1+\chi_{2N-1}^{-}\chi_{2N}^{-})(-1)^{F}_{\text{w}},\label{eq:TRR_w}
\end{align}
where the subscript RR stands for the Ramond-Ramond sector, following the continuum correspondence discussed in Ref. \cite{10.21468/SciPostPhys.16.3.064}. Note that we carefully redefine the phases so that $(T^{\pm}_{\text{RR}})^{L} = (T_{\text{RR}}^{\text{w}})^{L} = 1$ and no $N$ dependence in the symmetry algebraic structure (see below Sec. \ref{sec:anomaly_even_L_no_defect}).

When $L=2N+1$ is odd, the translation symmetries can be represented as
\begin{align}
    &T_{\text{NSR}}^{+} = e^{i\pi(xN+y)} \frac{e^{-i\frac{2N+1}{16}2\pi}}{2^{N}} (1-\chi_{1}^{+}\chi_{2}^{+}) (1-\chi_{2}^{+}\chi_{3}^{+}) \cdots (1-\chi_{2N}^{+}\chi_{2N+1}^{+}),\label{eq:TNSR_plus}\\
    &T_{\text{NSR}}^{-} = e^{-i\pi(xN+y)} \frac{e^{i\frac{2N+1}{16}2\pi}}{2^{N}} (1-\chi_{1}^{-}\chi_{2}^{-}) (1-\chi_{2}^{-}\chi_{3}^{-}) \cdots (1-\chi_{2N}^{-}\chi_{2N+1}^{-}),\label{eq:TNSR_minus}\\
    &T_{\text{NSR}}^{\text{w}} = \frac{1}{2^{2N}} (1-\chi_{1}^{+}\chi_{2}^{+})(1-\chi_{1}^{-}\chi_{2}^{-}) (1-\chi_{2}^{+}\chi_{3}^{+})(1-\chi_{2}^{-}\chi_{3}^{-}) \cdots (1-\chi_{2N}^{+}\chi_{2N+1}^{+})(1-\chi_{2N}^{-}\chi_{2N+1}^{-}),\label{eq:TNSR_w}
\end{align}
where we follow Ref. \cite{10.21468/SciPostPhys.16.3.064} to choose $x=-\frac{n_{1}}{2}$ and $y=\frac{3n_{1}}{4}+n_{2}$, with $n_{1},n_{2}\in\mathbb{Z}$. The subscript NS stands for the Neveu--Schwarz boundary condition in the continuum theory. The strong translation symmetries satisfy
\begin{align}
    \left(T_{\text{NSR}}^{\pm}\right)^{L} = e^{\pm i\frac{n}{16}2\pi},\quad
    n=6n_{1}+8n_{2}-1 \in 2\mathbb{Z}+1.
\end{align}
Although the phase is expressed as a sixteenth root of unity, allowed local phase redefinitions change $n$ by arbitrary even integers. Only its oddness is invariant, yielding a modulo-$2$ lattice obstruction.
This obstruction persists for the strong translations in open quantum systems, whereas the conjugate phases cancel for the weak translation symmetry, giving $(T_{\text{NSR}}^{\text{w}})^{L}=1$. We will discuss the modified translation symmetries later when we introduce the fermion-parity defect.

\subsection{Examples}\label{sec:jump_operator_examples}
Equation (\ref{eq:L0_collective_density_dephasing_jump_op}) satisfies both the strong fermion parity and strong translation symmetries. In this subsection, we provide several other examples of jump operators to illustrate situations in which the strong fermion parity or/and strong translation is/are broken. We emphasize that, for the anomaly analysis considered here, the explicit microscopic form of the Lindbladian is secondary---the symmetry algebra constrains the anomaly structure and the associated symmetry-enforced degeneracies of the Lindbladian spectrum, although quantitative features of the dynamics depend on the microscopic details. We hope that these examples can serve as starting points for exploring possible physical realizations of the anomaly-related phenomena.

\noindent\textbf{Strong fermion parity $(-1)^{F^\pm}$ is broken but strong translation $T^\pm$ remains preserved.} The strong fermion parity is broken, while the weak fermion parity holds, when the jump operator is fermion-parity odd. Microscopically, such a jump operator can arise through exchange of fermions or Bogoliubov quasiparticles with a fermionic reservoir\cite{PhysRevB.85.174533}. When the strong (one site) translation symmetry is further applied, the jump operator is restricted to, for example, the form of
\begin{align}
    L = \sum_{l} \chi_l.\label{eq:L_ex_linear_combination_of_chi}
\end{align}
A possible engineered realization may consist of a sufficiently short Majorana chain coherently coupled to a superconducting reservoir. Since the complex fermions are related to Majorana fermions through $c_j=(\chi_{2j-1}+i\chi_{2j})/2$, exchange of Bogoliubov quasiparticles can in principle generate fermion-parity-odd jumps involving linear combination of Majorana fermions. Realizing the uniform mode in Eq. (\ref{eq:L_ex_linear_combination_of_chi}), however, requires the reservoir to couple predominantly to a single collective mode of the chain, and this should be regarded as an engineered collective-coupling limit rather than a generic consequence of coupling a chain to a two-dimensional reservoir.

\noindent\textbf{Both strong fermion parity $(-1)^{F^\pm}$ and strong translation $T^\pm$ are broken}
When the spatial correlations of the fermionic reservoir are sufficiently short ranged, the effective dissipation can instead become local. In the ideal limit of independent and identical local bath channels, one may consider
\begin{align}
    L_m = \chi_m,
\end{align}
with $m=1,2,\cdots,L$. Each jump operator is fermion-parity odd and therefore breaks the strong fermion-parity symmetry. Translation permutes the jump operators, $L_m\rightarrow L_{m+1}$, so that the full Lindbladian remains weakly translation invariant, while strong translation is broken.


\noindent\textbf{Strong fermion parity $(-1)^{F^\pm}$ remains preserved while strong translation $T^\pm$ is broken}
The strong fermion parity symmetry holds when the jump terms are fermion-parity even. The requirement that only the weak translation symmetry is satisfied arises when each site couples to independent but identical bath. For instance,
\begin{align}
    L_m = \chi_m \chi_{m+1},
\end{align}
where $m=1,2,\cdots L$ and $\chi_{L+1}\sim\chi_1$. Similar to the case where both strong fermion parity and strong translation are broken, translation maps $L_m$ to $L_{m+1}$, so that, for identical bath, the full Lindbladian retains weak translation symmetry.

\section{Anomaly of Lindbladian Majorana Chain}\label{eq:anomaly_of_Lindbladian_Majorana_chain}
In this section, we apply the defect-based diagnosis of anomalies to a the Lindbladian Majorana chain. For closed systems, an anomaly can be detected by comparing how the symmetry algebra is realized in the ordinary sector and in sector with symmetry defects, since a defect implements a background symmetry flux, and an anomaly is precisely an obstruction to defining the symmetry consistently in all such background sectors. The same logic can also apply to the Liouville-space formulation of a Lindbladian system: one may introduce defect and no-defect sectors for the doubled space degrees of freedom and compare the corresponding algebras of strong and weak symmetries. A robust mismatch between these algebras, after allowing for reasonable phase redefinition, signals an anomaly. We remark, however, that this diagnostic does not exhaust all possible anomalies of Majorana chains. In particular, the modulo-$2$ lattice obstruction is more subtle and appears in the NSR and RNS sectors. We will show that the anomalies already visible in the closed Majorana chain survive in the Lindbladian problem as anomalies involving strong symmetries. More importantly, the open system setting allows additional anomaly structures involving weak symmetries, which have no direct counterpart in closed systems. We further show that, for the odd-length dissipative Majorana chain with $L=2N+1$, various fermion-parity-involved anomaly algebras can be realized. These anomalies are intrinsic to the Lindbladian formulation, in the sense that the relevant fermion-parity operations act naturally in Liouville space but do not arise as a inner automorphism in the corresponding closed Majorana chain.

\subsection{Even $L$ without Defect}\label{sec:anomaly_even_L_no_defect}
The fermion-parity symmetries, translation symmetries, and modular conjugation satisfy the following closed symmetry algebra,
\begin{align}
& \mathcal{J}^2=1,\label{eq:algebra_RR_start}\\
& \left[(-1)^{F^\pm}\right]^2 = \left[(-1)^{F^\text{w}}\right]^2=1,\\
& \left[T_{\text{RR}}^\pm\right]^L = \left[ T_{\text{RR}}^{\text{w}}\right]^L = 1,\label{eq:TRR_to_L_equals_1}\\
& (-1)^{F^{+}} (-1)^{F^{-}} = (-1)^{F^\text{w}},\\
& T_{\text{RR}}^{+} T_{\text{RR}}^{-} = i T_{\text{RR}}^{\text{w}},\label{eq:TRR_plus_TRR_minus_equals_i_TRR_w}\\
& \mathcal{J} (-1)^{F^{+}} \mathcal{J}^{-1} = (-1)^{F^{-}},\\
& \mathcal{J} T_{\text{RR}}^{+} \mathcal{J}^{-1} = T_{\text{RR}}^{-},\\
& (-1)^{F^{+}} T_{\text{RR}}^{+} = - T_{\text{RR}}^{+} (-1)^{F^{+}},\label{eq:strong_parity_and_T_anticommute}\\
& (-1)^{F^{+}} T_{\text{RR}}^{-} = T_{\text{RR}}^{-} (-1)^{F^{+}},\\
& T_{\text{RR}}^{+} T_{\text{RR}}^{-} = - T_{\text{RR}}^{-} T_{\text{RR}}^{+}.\label{eq:algebra_RR_end}
\end{align}
The minus signs, which cannot be eliminated by phase redefinition, clearly signal an anomaly in the quantum-mechanical sense. The phase factor $i$ cannot be removed unless one abandons the condition $(T_{\text{RR}}^{\text{w}})^{L}=1$. Other commutation relations such as $(-1)^{F^+}(-1)^{F^-}=(-1)^{F^-}(-1)^{F^+}$ and $(-1)^{F^-} T_{\text{RR}}^- = - T_{\text{RR}}^- (-1)^{F^-}$ can be derived from the minimal generating set above. In particular, we have 
\begin{align}
(-1)^{F^\pm} T_{\text{RR}}^\text{w} = - T_{\text{RR}}^\text{w} (-1)^{F^\pm}
\label{eq:strong_fermion_parities_and_weak_translation_anticommute_RR}
\end{align}
This relation is one of our main results: even when the strong translation symmetries are broken, the surviving strong fermion-parity symmetries and the weak translation symmetry still form a projective symmetry algebra, manifesting the existence of anomaly. Since the corresponding operators are symmetries of the Lindbladian, their anticommutation pairs Liouvillian eigenmodes with opposite fermion-parity quantum numbers and, in particular, enforce degeneracy of the steady-state subspace. Because this projective relation involves both weak and strong symmetries, it is peculiar to open systems. As we will see in Sec. \ref{sec:anomaly_RR}, the same anomaly can be diagnosed from the twisted sector, where it appears as an obstruction to consistently obtaining the same symmetry algebra in the presence of a weak fermion-parity defect.

A similar situation arises when the strong fermion parity symmetries $(-1)^{F^\pm}$ are broken, while the weak fermion parity $(-1)^{F^\text{w}}$ and the strong translation symmetries $T_\text{RR}^\pm$ persists. In this case, we have the following anomalous relation:
\begin{align}
(-1)^{F^\text{w}} T_\text{RR}^\pm = - T_\text{RR}^\pm (-1)^{F^\text{w}}\label{eq:weak_fermion_parities_and_strong_translation_anticommute_RR}
\end{align}
which again constrains the steady-state degeneracy and is intrinsic to open quantum systems. As we will see in Sec. \ref{sec:anomaly_RR}, the corresponding twisted-sector obstruction takes a different form: the strong translation symmetries cannot be extended to the twisted sector when the strong fermion parity symmetries are absent.

\subsection{Even $L$ with Weak Fermion Parity Defect \label{sec:anomaly_RR}}
When a weak fermion parity defect is introduced, the symmetries that do not commute with $(-1)^{F^\text{w}}$ must be modified. In particular, we require the translation symmetries to follow
\begin{align}
    T^{\pm}_\mathsf{G}:&\quad \chi_{l}^{\pm}\rightarrow\chi_{l+1}^{\pm}, \quad \chi_{l}^{\mp}\rightarrow\chi_{l}^{\mp}\quad \text{for }l=1,2,\cdots,L-1,\label{eq:translation_w_defect_rule_1}\\
    &\quad \chi_{L}^{\pm}\rightarrow-\chi_{1}^{\pm},\quad \chi_{L}^{\mp}\rightarrow\chi_{L}^{\mp},\\
    T_{\mathsf{G}}^\text{w}:&\quad \chi_{l}^{\pm}\rightarrow\chi_{l+1}^{\pm}\quad \text{for }l=1,2,\cdots,L-1,\\
    &\quad \chi_{L}^{\pm}\rightarrow-\chi_{1}^{\pm},\label{eq:translation_w_defect_rule_4}
\end{align}
where the subscript $\mathsf{G}$ denotes the presence of the defect and the minus signs are designed to compensate for the minus signs from the fermion-parity defect.

After a suitable phase redefinition, these translation symmetries can be realized as
\begin{align}
    &T_{\text{NSNS}}^{+}=\frac{  e^{-i\frac{N}{4}\pi}  }{  2^{\frac{2N-1}{2}}  }
    (1-\chi_{1}^{+}\chi_{2}^{+})(1-\chi_{2}^{+}\chi_{3}^{+})\cdots
    (1-\chi_{2N-1}^{+}\chi_{2N}^{+}),\label{eq:TNSNS_plus}\\
    &T_{\text{NSNS}}^{-}=\frac{  e^{i\frac{N}{4}\pi}  }{  2^{\frac{2N-1}{2}}  }
    (1-\chi_{1}^{-}\chi_{2}^{-})(1-\chi_{2}^{-}\chi_{3}^{-})\cdots
    (1-\chi_{2N-1}^{-}\chi_{2N}^{-}),\label{eq:TNSNS_minus}\\
    &T_{\text{NSNS}}^{\text{w}} = \frac{  1  }{  2^{2N-1}  }
    (1-\chi_{1}^{+}\chi_{2}^{+})(1-\chi_{1}^{-}\chi_{2}^{-}) (1-\chi_{2}^{+}\chi_{3}^{+}) (1-\chi_{2}^{-}\chi_{3}^{-})\cdots
    (1-\chi_{2N-1}^{+}\chi_{2N}^{+})(1-\chi_{2N-1}^{-}\chi_{2N}^{-}),\label{eq:TNSNS_w}
\end{align}
where the subscript denotes the anti-periodic boundary condition (NS) for both left- and right- movers in Majorana CFT.

Together with $(-1)^{F^\text{w}}$, $(-1)^{F^{\pm}}$, and $\mathcal{J}$, they form the following algebra:
\begin{align}
& \mathcal{J}^2=1,\label{eq:algebra_NSNS_start}\\
& \left[(-1)^{F^\pm}\right]^2 = \left[(-1)^{F^\text{w}}\right]^2=1,\\
& \left[T_{\text{NSNS}}^\pm\right]^L = (-1)^{F^\pm},\label{eq:NSNS_Tpm_to_L_equals_to_parity_pm}\\
& \left[ T_{\text{NSNS}}^{\text{w}}\right]^L = (-1)^{F^\text{w}},\\
& (-1)^{F^{+}} (-1)^{F^{-}} = (-1)^{F^\text{w}},\\
& T_{\text{NSNS}}^{+} T_{\text{NSNS}}^{-} = T_{\text{NSNS}}^{\text{w}},\\
& \mathcal{J} (-1)^{F^{+}} \mathcal{J}^{-1} = (-1)^{F^{-}},\\
& \mathcal{J} T_{\text{NSNS}}^{+} \mathcal{J}^{-1} = T_{\text{NSNS}}^{-},\\
& (-1)^{F^{+}} T_{\text{NSNS}}^{+} = T_{\text{NSNS}}^{+} (-1)^{F^{+}},\\
& (-1)^{F^{+}} T_{\text{NSNS}}^{-} = T_{\text{NSNS}}^{-} (-1)^{F^{+}},\\
& T_{\text{NSNS}}^{+} T_{\text{NSNS}}^{-} = T_{\text{NSNS}}^{-} T_{\text{NSNS}}^{+}.\label{eq:algebra_NSNS_end}
\end{align}
More importantly, when the strong translation symmetries are broken while the strong fermion parity symmetries remain preserved,
\begin{align}
    (-1)^{F^\pm} T_{\text{NSNS}}^\text{w} = T_{\text{NSNS}}^\text{w} (-1)^{F^\pm}. \label{eq:wTransNSNS_sFP}
\end{align}
The commutation relation in Eq. \eqref{eq:wTransNSNS_sFP} contrasts with the
anticommutation relation in Eq. \eqref{eq:strong_fermion_parities_and_weak_translation_anticommute_RR} in the untwisted sector.
Moreover, Eq. \eqref{eq:NSNS_Tpm_to_L_equals_to_parity_pm} constrains the realization of strong
translation symmetry in the weak-fermion-parity-twisted sector.
Let $\mathcal{L}_\mathsf{G}$ denote the twisted Lindbladian. Since
$(T^\pm_{\mathrm{NSNS}})^L = (-1)^{F^\pm}$, we have
$[T^\pm_{\mathrm{NSNS}},\mathcal{L}_\mathsf{G}]=0
\quad\Longrightarrow\quad
[(-1)^{F^\pm},\mathcal{L}_\mathsf{G}]=0$.
Consequently, if $\mathcal{L}_\mathsf{G}$ explicitly breaks a strong
fermion-parity symmetry, the corresponding twisted translation
$T^\pm_{\mathrm{NSNS}}$ defined in Eqs. \eqref{eq:TNSNS_plus} and \eqref{eq:TNSNS_minus}
cannot be a symmetry of $\mathcal{L}_\mathsf{G}$.

\subsection{Odd $L$ without Defect\label{sec:anomaly_odd_L}}
When $L=2N+1$ is odd, the fermion-parity symmetries, translation symmetries, and modular conjugation satisfy the following symmetry algebra,
\begin{align}
& \mathcal{J}^2=1,\label{eq:algebra_NSR_start}\\
& \left[(-1)^{F^\pm}\right]^2 = \left[(-1)^{F^\text{w}}\right]^2=1,\\
& \left[T_{\text{NSR}}^\pm\right]^L = e^{\pm i\frac{n}{16}2\pi},\\
& \left[ T_{\text{NSR}}^{\text{w}}\right]^L = 1,\\
& (-1)^{F^{+}} (-1)^{F^{-}} = i (-1)^{F^\text{w}},\label{eq:relation_bt_strong_weak_fermion_parities_anomaly}\\
& T_{\text{NSR}}^{+} T_{\text{NSR}}^{-} = T_{\text{NSR}}^{\text{w}},\\
& \mathcal{J} (-1)^{F^{+}} \mathcal{J}^{-1} = (-1)^{F^{-}},\\
& \mathcal{J} T_{\text{NSR}}^{+} \mathcal{J}^{-1} = T_{\text{NSR}}^{-},\\
& (-1)^{F^{+}} T_{\text{NSR}}^{+} = T_{\text{NSR}}^{+} (-1)^{F^{+}},\\
& (-1)^{F^{+}} T_{\text{NSR}}^{-} = T_{\text{NSR}}^{-} (-1)^{F^{+}},\\
& T_{\text{NSR}}^{+} T_{\text{NSR}}^{-} = T_{\text{NSR}}^{-} T_{\text{NSR}}^{+},\label{eq:algebra_NSR_end}
\end{align}
where, as described in Sec. \ref{sec:translation_symm}, $n \in 2\mathbb{Z}+1$ signals a modulo-$2$ lattice obstruction. In addition, when the strong fermion parity symmetries are satisfied, the central extension in Eq. (\ref{eq:relation_bt_strong_weak_fermion_parities_anomaly}) still indicates a quantum-mechanical anomaly.

\subsection{Odd $L$ with Weak Fermion Parity Defect}\label{sec:anomaly_RNS_odd_with_defect}
Unlike in the closed Majorana system, the dissipative Majorana chain allows a weak fermion party defect even when $L=2N+1$ is odd. The requirements for the translation symmetries, Eqs. (\ref{eq:translation_w_defect_rule_1})-(\ref{eq:translation_w_defect_rule_4}), can be satisfied through
\begin{align}
    &T^{+}_\text{RNS} = e^{i\pi(x'N+y')} \frac{1}{2^{\frac{L-1}{2}}}
    \chi_{1}^{+}
    (1-\chi_{1}^{+}\chi_{2}^{+})
    (1-\chi_{2}^{+}\chi_{3}^{+}) \cdots
    (1-\chi_{L-1}^{+}\chi_{L}^{+})
    (-1)^{F^\text{w}},\label{eq:TRNS_plus}\\
    &T^{-}_\text{RNS} = e^{-i\pi(x'N+y')} \frac{1}{2^{\frac{L-1}{2}}}
    \chi_{1}^{-}
    (1-\chi_{1}^{-}\chi_{2}^{-})
    (1-\chi_{2}^{-}\chi_{3}^{-}) \cdots
    (1-\chi_{L-1}^{-}\chi_{L}^{-})
    (-1)^{F^\text{w}},\label{eq:TRNS_minus}\\
    &T^{\text{w}}_\text{RNS} = \frac{i}{2^{L-1}}
    \chi_{1}^{+}\chi_{1}^{-}
    (1-\chi_{1}^{+}\chi_{2}^{+})
    (1-\chi_{1}^{-}\chi_{2}^{-})
    (1-\chi_{2}^{+}\chi_{3}^{+})
    (1-\chi_{2}^{-}\chi_{3}^{-}) \cdots
    (1-\chi_{L-1}^{+}\chi_{L}^{+})
    (1-\chi_{L-1}^{-}\chi_{L}^{-}),\label{eq:TRNS_w}
\end{align}
where $x' = n_{1}'/2 - 1/4$ and $y' = - 3n_{1}'/4 + n_{2}' + 5/8$ with $n_{1}',n_{2}'\in\mathbb{Z}$. The strong symmetries, similar to $T_{\text{NSR}}^{\pm}$, satisfy
\begin{align}
    (T_\text{RNS}^{\pm})^{L} = e^{\pm i \frac{n'}{16} 2\pi} (-1)^{F^{\pm}},\quad
    n' = - 6n_{1}' + 8n_{2}' + 9 \in 2\mathbb{Z}+1,
\end{align}
where the phase retains the modulo-$2$ obstruction associated with the oddness of $n'$. Furthermore, the translation around the defect pumps a strong fermion parity charge, indicating the existence of a mixed anomaly.

The surprising part comes in the algebraic structure: together with the fermion parity symmetries and modular conjugation, they satisfy
\begin{align}
&\mathcal{J}^2 = 1,\label{eq:algebra_RNS_start}\\
& \left[(-1)^{F^\pm}\right]^2 = \left[(-1)^{F^\text{w}}\right]^2=1,\\
& \left[T_\text{RNS}^\pm\right]^L = e^{\pm i\frac{n'}{16}2\pi} (-1)^{F^\pm},\\
& \left[ T_\text{RNS}^{\text{w}}\right]^L = (-1)^{F^\text{w}},\\
& (-1)^{F^{+}} (-1)^{F^{-}} = i (-1)^{F^\text{w}},\\
& T_\text{RNS}^{+} T_\text{RNS}^{-} = i T_\text{RNS}^{\text{w}},\\
& \mathcal{J} (-1)^{F^{+}} \mathcal{J}^{-1} = (-1)^{F^{-}},\\
& \mathcal{J} T_\text{RNS}^{+} \mathcal{J}^{-1} = T_\text{RNS}^{-},\\
& (-1)^{F^{+}} T_\text{RNS}^{+} = T_\text{RNS}^{+} (-1)^{F^{+}},\\
& (-1)^{F^{+}} T_\text{RNS}^{-} = - T_\text{RNS}^{-} (-1)^{F^{+}},\\
& T_\text{RNS}^{+} T_\text{RNS}^{-} = - T_\text{RNS}^{-} T_\text{RNS}^{+},\label{eq:algebra_RNS_end}
\end{align}
The symmetry operators obey a projective algebra whose commutation relations differ from those in the untwisted sector, signaling the existence of lattice anomaly intrinsic to the open quantum systems.


\section{IR Limit of Dissipative Majorana Chain}\label{sec:IR_limit_of_dissipative_Majorana_chain}
In this section, we establish the IR correspondence to the previous lattice results. In particular, we explain how microscopic lattice translation symmetries emanates to continuum translation symmetries dressed by chiral fermion parities. For the RR, NSNS, NSR, and RNS theories, the corresponding dictionaries take the following form,
\begin{align}
    &T^+ \sim (-1)^{F_\text{L}^+} e^{i 2\pi P^+ / L},\label{eq:dictionary_plus}\\
    &T^- \sim (-1)^{F_\text{L}^-} e^{-i 2\pi P^- / L},\label{eq:dictionary_minus}\\
    &T^{\text{w}} \sim (-1)^{F_\text{L}^+} (-1)^{F_\text{L}^-} e^{i 2\pi P^+ / L} e^{-i 2\pi P^- / L},\label{eq:dictionary_w}
\end{align}
where $T^\pm$ and $T^\text{w}$ denote strong and weak lattice translation symmetries, respectively. The operators $P^\pm$ are the continuum momentum operators, and $(-1)^{F_\text{L}^\pm}$ are the emanant strong chiral fermion parities of the left-moving Majorana modes. The symbol $\sim$ indicates equality up to a sector-dependent phase factor. Using these dictionaries, we then derive the continuum symmetry algebra from the exact lattice symmetries. This allows us to identify which anomalous features of the open-system symmetry algebra persist in the long-wavelength description.

We remark that the IR limit considered here is taken with respect to the Hamiltonian part of the Lindbladian. It should not be confused with the slow-mode limit of the full Liouvillian. The purpose of the Hamiltonian IR limit is to provide a continuum operator algebra in which the exact microscopic symmetries of the Lindbladian can be represented. When the system-bath coupling is weak, irrelevant, or treated as a symmetry-preserving perturbation of the critical Hamiltonian, this continuum description also serves as the natural starting point for the open-system dynamics. For a generic dissipator, however, the true long-time behavior is governed by the full Lindbladian and may differ from the Hamiltonian critical theory.

Let us first recall the closed Majorana Hamiltonian in the IR limit mentioned in Sec. \ref{sec:Majorana_chain_and_sym_in_Hamiltonian_systems},
\begin{align}
    H_{\text{Maj}} = i \int dx\, \left( \chiL \partial_x \chiL - \chiR \partial_x \chiR \right),
\end{align}
which, after the mode expansion
\begin{align}
    &\chiL(t,x) = \frac{1}{\sqrt{4\pi}} \sum_{r\in\mathbb{Z}+\nuL} \chi_{\text{L},r} e^{-ir(t+x)},\\
    &\chiR(t,x) = \frac{1}{\sqrt{4\pi}} \sum_{r\in\mathbb{Z}+\nuR} \chi_{\text{R},r} e^{-ir(t-x)}
\end{align}
takes the form
\begin{align}
    H_{\text{Maj}} = \sum_{r=1-\nuL}^{\infty} r \chi_{\text{L},-r} \chi_{\text{L},r}
    + \sum_{r=1-\nuR}^{\infty} r \chi_{\text{R},-r} \chi_{\text{R},r} + \text{const.}
\end{align}
Here $\nuL$ and $\nuR$ specify the spin structures of the left- and right-moving Majorana modes. In the Ramond (R) sector the corresponding modes are integer-moded ($\nu=0$), while in the Neveu--Schwarz (NS) sector they are half-integer-moded ($\nu=\frac{1}{2}$). This continuum Hamiltonian $H_\text{Maj}$ will serve as a reference Hamiltonian when we identify the lattice modes with the continuum modes.

We then introduce the corresponding ket- and bra-sector fields using Eqs. (\ref{eq:def_A_plus}) and (\ref{eq:def_A_minus}):
\begin{align}
    &\chiL^+(t,x) = \frac{1}{\sqrt{4\pi}} \sum_{r\in\mathbb{Z}+\nuL} \chi^+_{\text{L},r} e^{-ir(t+x)},\\
    &\chiR^+(t,x) = \frac{1}{\sqrt{4\pi}} \sum_{r\in\mathbb{Z}+\nuR} \chi^+_{\text{R},r} e^{-ir(t-x)},\\
    &\chiL^-(t,x) = \frac{1}{\sqrt{4\pi}} \sum_{r\in\mathbb{Z}+\nuL} \chi^-_{\text{L},r} e^{ir(t+x)},\\
    &\chiR^-(t,x) = \frac{1}{\sqrt{4\pi}} \sum_{r\in\mathbb{Z}+\nuR} \chi^-_{\text{R},r} e^{ir(t-x)},
\end{align}
where the ket- and bra-sector Majorana fields are related by the modular conjugation $\mathcal{J}$. Since $\mathcal{J}$ is antiunitary, it complex-conjugates the phase factors in the mode expansions. The corresponding Hamiltonians are
\begin{align}    
    &H_{\text{Maj}}^+ = \sum_{r=1-\nuL}^{\infty} r \chi^+_{\text{L},-r} \chi^+_{\text{L},r}
    + \sum_{r=1-\nuR}^{\infty} r \chi^+_{\text{R},-r} \chi^+_{\text{R},r} + \text{const.},\label{eq:H_Maj_plus}\\
    &H_{\text{Maj}}^- = \sum_{r=1-\nuL}^{\infty} r \chi^-_{\text{L},-r} \chi^-_{\text{L},r}
    + \sum_{r=1-\nuR}^{\infty} r \chi^-_{\text{R},-r} \chi^-_{\text{R},r} + \text{const.}.\label{eq:H_Maj_minus}
\end{align}
They are associated by modular conjugate as well, and will serve as the CFT reference Hamiltonians to which we compare the IR limit of the lattice theory. We now discuss the four spin sectors separately.

\subsection{RR Sector}
In this subsection, we focus on the Hamiltonian IR limit of the no-defect sector with even system size $L=2N$. We begin with the Majorana chain Hamiltonain in the ket and bra representatives:
\begin{align}
    H_0^\pm &= \pm \frac{i}{2}\sum_{l=1}^{L-1} \chi_{l+1}^\pm \chi_l^\pm \pm \frac{i}{2} \chi_1^\pm \chi_L^\pm.\label{eq:tmpt_1}
\end{align}
We take the following mode expansion,
\begin{align}
    &\chi_l^{\pm} = \sqrt{\frac{2}{L}} \sum_{k=1}^L e^{ \pm i 2\pi kl/L} d_k^{\pm},\label{eq:mode_expansion_chi_RR}
\end{align}
with inverse
\begin{align}
    d_k^\pm = \frac{1}{\sqrt{2L}} \sum_{l=1}^L e^{\mp i 2\pi kl/L} \chi_l^\pm, \label{eq:d_k_pm_inverse_RR}
\end{align}
where $k$ and $l$ are understood modulo $L$. Since the Majorana fermions are Hermitian, the Fourier modes obey
\begin{align}
    d_k^{\pm} = \left(d_{-k}^{\pm}\right)^{\dagger}, \quad
    d_0^{\pm} = \left(d_0^{\pm}\right)^\dagger, \quad
    d_N^{\pm} = \left(d_N^{\pm}\right)^\dagger,
\end{align}
and the canonical anticommutation relations,
\begin{align}
    \left\{ d_k^s, d_{k'}^{s'} \right\} = \delta_{s,s'}\delta_{k,-k'}.
\end{align}
Substituting the mode expansion into Eq. (\ref{eq:tmpt_1}), we obtain
\begin{align}
    H_0^{\pm} &= \pm i \sum_{k=0}^{L-1} e^{\pm i\pi k/N} d_k^\pm d_{-k}^\pm = 2 \sum_{k=1}^{N-1} \sin{\left(\frac{\pi k}{N}\right)} \left(d_k^\pm\right)^\dagger d_k^\pm + \text{const.}
\end{align}
In the large-$L$ limit, the low-energy modes are localized near $k=0$ and $k=N$. Keeping modes with $1\ll\Lambda\ll N$, we find
\begin{align}
    H_0^{\pm} &\simeq \frac{2\pi}{N} \sum_{k=1}^{\Lambda} k \left(d_k^\pm\right)^\dagger d_k^\pm
    + \frac{2\pi}{N} \sum_{k=1}^{\Lambda} k \left(d_{N-k}^\pm\right)^\dagger d_{N-k}^\pm + \text{const.}
\end{align}
Comparing this expression to the continuum Hamiltonians in (\ref{eq:H_Maj_plus}) and (\ref{eq:H_Maj_minus}), we identify the continuum right- and left-moving modes as
\begin{align}
    &\chi_{\text{R},k}^\pm = d_k^\pm,\label{eq:identification_RR_1}\\
    &\chi_{\text{L},k}^\pm = d_{N-k}^\pm.\label{eq:identification_RR_2}
\end{align}

We now examine the actions of lattice symmetries to the low-energy modes. The fermion parities in Eqs. (\ref{eq:strong_fermion_parity_plus_even}) and (\ref{eq:strong_fermion_parity_minus_even}) act on $d_k^\pm$ as
\begin{align}
    &(-1)^{F^+} d_k^\pm (-1)^{F^+} = \mp d_k^\pm,\\
    &(-1)^{F^-} d_k^\pm (-1)^{F^-} = \pm d_k^\pm,\\
    &(-1)^{F^\text{w}} d_k^\pm (-1)^{F^\text{w}} = -d_k^\pm.
\end{align}
The translation symmetries in Eqs. (\ref{eq:TRR_plus})-(\ref{eq:TRR_w}) act as
\begin{align}
    &T_{\text{RR}}^+ d_k^+ \left(T_{\text{RR}}^+\right)^{-1} = e^{i\pi k/N} d_k^+, \quad
    T_{\text{RR}}^+ d_k^- \left(T_{\text{RR}}^+\right)^{-1} = d_k^-,\\
    &T_{\text{RR}}^- d_k^+ \left(T_{\text{RR}}^-\right)^{-1} = d_k^+, \quad
    T_{\text{RR}}^- d_k^- \left(T_{\text{RR}}^-\right)^{-1} = e^{-i\pi k/N} d_k^-,\\
    &T_{\text{RR}}^\text{w} d_k^+ \left(T_{\text{RR}}^\text{w}\right)^{-1} = e^{i\pi k/N} d_k^+, \quad
    T_{\text{RR}}^\text{w} d_k^- \left(T_{\text{RR}}^\text{w}\right)^{-1} = e^{-i\pi k/N} d_k^-.
\end{align}
Using the identifications (\ref{eq:identification_RR_1}) and (\ref{eq:identification_RR_2}), these transformations become
\begin{align}
    &T_{\text{RR}}^+:\quad 
    \chi_{\text{R},k}^+ \rightarrow e^{i2\pi k/L} \chi_{\text{R},k}^+, \quad
    \chi_{\text{L},k}^+ \rightarrow - e^{-i2\pi k/L} \chi_{\text{L},k}^+, \quad
    \chi_{\text{R},k}^- \rightarrow \chi_{\text{R},k}^-, \quad
    \chi_{\text{L},k}^- \rightarrow \chi_{\text{L},k}^-,\\
    &T_{\text{RR}}^-:\quad
    \chi_{\text{R},k}^+ \rightarrow \chi_{\text{R},k}^+,\quad
    \chi_{\text{L},k}^+ \rightarrow \chi_{\text{L},k}^+,\quad
    \chi_{\text{R},k}^- \rightarrow e^{-i2\pi k/L} \chi_{\text{R},k}^-, \quad
    \chi_{\text{L},k}^- \rightarrow -e^{i2\pi k/L} \chi_{\text{L},k}^-,\\
    &T_{\text{RR}}^\text{w}:\quad
    \chi_{\text{R},k}^+ \rightarrow e^{i2\pi k/L} \chi_{\text{R},k}^+, \quad
    \chi_{\text{L},k}^+ \rightarrow - e^{-i2\pi k/L} \chi_{\text{L},k}^+, \quad
    \chi_{\text{R},k}^- \rightarrow e^{-i2\pi k/L} \chi_{\text{R},k}^-, \quad
    \chi_{\text{L},k}^- \rightarrow -e^{i2\pi k/L} \chi_{\text{L},k}^-,
\end{align}
from which we can infer the following RR-sector dictionaries:
\begin{align}
    &T_\text{RR}^+ = (-1)^{F_\text{L}^+} e^{i2\pi P^+/L},\label{eq:relation_TRRp_fermion_parity_L_plus_momentum}\\
    &T_\text{RR}^- = (-1)^{F_\text{L}^-} e^{-i2\pi P^-/L},\label{eq:relation_TRRm_fermion_parity_L_minus_momentum}\\
    &T_\text{RR}^\text{w} = -i  (-1)^{F_\text{L}^+} (-1)^{F_\text{L}^-} e^{i2\pi P^+/L} e^{-i2\pi P^-/L},\label{eq:relation_TRRw_fermion_parity_L_pm_momentum}
\end{align}
where $(-1)^{F_\text{L}^\pm}$ operate on the continuum modes as
\begin{align}
    &(-1)^{F_\text{L}^+}:\quad \chiLk^+\rightarrow-\chiLk^+, \quad
    \chiRk^+\rightarrow\chiRk^+,\quad
    \chi_\text{L/R,k}^-\rightarrow\chi_\text{L/R,k}^-\label{eq:fermion_parity_L_plus}\\
    &(-1)^{F_\text{L}^-}:\quad \chiLk^-\rightarrow-\chiLk^-, \quad
    \chiRk^-\rightarrow\chiRk^-,\quad
    \chi_\text{L/R,k}^+\rightarrow\chi_\text{L/R,k}^+\label{eq:fermion_parity_L_minus}
\end{align}
while the continuum translation generated by $P^\pm$ act as
\begin{align}
    &e^{i2\pi P^+/L}:\quad \chiLk^+\rightarrow e^{-i2\pi k/L}\chiLk^+, \quad
    \chiRk^+\rightarrow e^{i2\pi k/L}\chiRk^+,\quad
    \chi_\text{L/R,k}^-\rightarrow\chi_\text{L/R,k}^-,\\
    &e^{-i2\pi P^-/L}:\quad \chiLk^-\rightarrow e^{i2\pi k/L}\chiLk^-, \quad
    \chiRk^-\rightarrow e^{-i2\pi k/L}\chiRk^-,\quad
    \chi_\text{L/R,k}^+\rightarrow\chi_\text{L/R,k}^+.
\end{align}
Strictly speaking, $T_\text{RR}^\pm$ and $T_\text{RR}^\text{w}$ determine only the operator structure of the dictionaries, and the overall phases are not determined yet. They can be chosen by requiring consistency with the exact lattice symmetry algebra, i.e. Eqs. (\ref{eq:TRR_to_L_equals_1}) and (\ref{eq:TRR_plus_TRR_minus_equals_i_TRR_w}). We remark that, in the dissipative Majorana chain, the translation symmetries emanate to the continuum translations dressed by the left fermion parities.

There is also a subtlety regarding the strong chiral fermion parities. The operator $(-1)^{F_\text{L}^+}$ should not be regarded as a naive ket-lift from the Hilbert space operator $(-1)^{F_\text{L}}$. In the RR sector, the Hilbert-space operator $(-1)^{F_\text{L}}$ is represented, up to convention, by the Majorana zero mode $\chiRz$ in the zero-mode subspace. If one naively lifts this operator to the ket sector $(-1)^{F_\text{L}^+}\sim\chiRz^+$, according to Eq. (\ref{eq:relation_TRRp_fermion_parity_L_plus_momentum}) one would obtain $T_\text{RR}^+\sim\chiRz^+$, and acting this $T_\text{RR}^+$ on minus-sector $\chi_\text{R/L,k}^-$ gives an extra minus sign. This contradicts the transformation rule in Eq. (\ref{eq:relation_TRRp_fermion_parity_L_plus_momentum}), where $(-1)^{F_\text{L}^+}$ leaves the minus sector invariant. Therefore, $(-1)^{F_\text{L}^+}$ must include an additional $(-1)^{F^-}$ to compensate the minus sign. Similarly, $(-1)^{F_\text{L}^-}$ cannot be a naive bra-lift from $(-1)^{F_\text{L}}$---it must include $(-1)^{F^+}$.

Using these dictionaries together with the exact lattice algebra in Eqs. (\ref{eq:algebra_RR_start})-(\ref{eq:algebra_RR_end}), we obtain the following continuum symmetry algebra in the Hamiltonian IR limit:
\begin{align}
& e^{\pm i 2\pi P^\pm} = 1,\label{eq:RR_compatible_relation}\\
& (-1)^{F^{+}} (-1)^{F_\text{L}^+} = - (-1)^{F_\text{L}^+} (-1)^{F^{+}},\\
& (-1)^{F^{+}} (-1)^{F_\text{L}^-} = (-1)^{F_\text{L}^-} (-1)^{F^{+}},\\
&(-1)^{F_\text{L}^+} (-1)^{F_\text{L}^-} = - (-1)^{F_\text{L}^-} (-1)^{F_\text{L}^+}.
\end{align}
Here we have displayed the relations relevant for the translation dictionaries; the remaining fermion-parity relations are inherited directly from the lattice theory. Equation (\ref{eq:RR_compatible_relation}) is consistent with the RR theory. The minus signs reproduce the anomalous signs already present in the microscopic lattice algebra. Thus the anomaly survives in the Hamiltonian IR description.

We emphasize that this conclusion concerns the Hamiltonian IR limit. For different choices of jump operators, two additional effects may occur: (i) some symmetries might be explicitly broken, reducing the symmetry group and its algebra, and (ii) the Majorana zero modes may be damped by the dissipative dynamics and therefore need not survive as the zero modes of the full Lindbladian. These questions depend on the detailed dissipator and will be left for future work.

As a final remark, the jump operators introduced earlier can take simple forms in this low-energy limit. For instance, the collective Majorana jump in Eq. (\ref{eq:L_ex_linear_combination_of_chi}), which preserves weak fermion parity and strong translation but breaks strong fermion parity, can be written as
\begin{align}
    L=\sum_{l=1}^L\chi_l = \sqrt{2L}d_0,
\end{align}
and therefore becomes proportional $\chiRz$ in the continuum. Absorbing the normalization into an effective coupling strength $\gamma$, we write the continuum jump as $\sqrt{\gamma} \chiRz$.

Using the doubled space convention of Eq. (\ref{eq:Lindbladian_in_doubled_Hilbert_space}), the continuum Lindbladian is
\begin{align}
    \mathcal{L} = -i \left(H_{\text{Maj}}^+ - H_{\text{Maj}}^-\right) - i \gamma \chiRz^+ \chiRz^- - \frac{\gamma}{2},
\end{align}
The dissipative term commutes with weak fermion parity $(-1)^{F^\text{w}}$, but not with either strong fermion parity $(-1)^{F^\pm}$. It also commutes with the momenta $P^\pm$ and the strong left-chiral fermion parities $(-1)^{F_\text{L}^\pm}$, consistently with the dictionaries in Eqs. (\ref{eq:dictionary_plus}) and (\ref{eq:dictionary_minus}). Thus, the continuum Lindbladian preserves weak fermion parity and the continuum images of the strong lattice translations.

In Sec. \ref{sec:anomaly_even_L_no_defect}, the lattice analysis already establishes spectral degeneracy in the presence of anomaly. We now illustrate its continuum counterpart. First, we focus on the zero mode contribution. Using the following Jordan--Wigner representation,
\begin{align}
    &\chiRz^+ = \frac{1}{\sqrt{2}} \sigma^x \otimes I, \qquad
    \chiLz^+ = \frac{1}{\sqrt{2}} \sigma^y \otimes I,\\
    &\chiRz^- = \frac{1}{\sqrt{2}} \sigma^z \otimes \sigma^x, \qquad
    \chiLz^- = \frac{1}{\sqrt{2}} \sigma^z \otimes \sigma^y,
\end{align}
the zero-mode contribution to the Lindbladian is
\begin{align}
    \mathcal{L}_0 = -i \gamma \chiRz^+ \chiRz^- - \frac{\gamma}{2}
    = -\frac{\gamma}{2} (\sigma^y \otimes \sigma^x + I \otimes I).
\end{align}
Its eigenvalues are
\begin{align}
    \lambda = \{0,0,-\gamma,-\gamma\}.
\end{align}
The twofold degeneracy reflects the anticommutation relation between the weak fermion parity and the strong left-chiral fermion parity, both of which commute with the Lindbladian. Second, restoring the nonzero modes, i.e. $H_\text{Maj}$, adds only imaginary energy differences to the eigenvalues, producing phase oscillations without lifting this degeneracy.

One way to break this anomaly-protected degeneracy is to include the generic mass term,
\begin{align}
    H_m = i m \chiRz \chiLz,
\end{align}
which preserves the strong fermion parity but breaks the left-chiral fermion parity of the total Lindbladian. The zero-mode Lindbladian becomes
\begin{align}
    \mathcal{L}_0 = i \frac{m}{2} (\sigma^z \otimes I + I \otimes \sigma^z)
    -\frac{\gamma}{2} (\sigma^y \otimes \sigma^x + I \otimes I)
\end{align}
with eigenvalues
\begin{align}
    \lambda = \left\{0, -\gamma, -\frac{\gamma}{2} \pm \sqrt{\frac{\gamma^2}{4}-m^2} \right\},
\end{align}
and the degeneracies are lifted.

\subsection{NSNS Sector}
In the NSNS sector, we focus on the Majorana chain Hamiltonian with the defect, namely Eq. (\ref{eq:H_G}). The corresponding ket- and bra-sector representatives are:
\begin{align}
    H_\mathsf{G}^\pm &= \pm \frac{i}{2}\sum_{l=1}^{L-1} \chi_{l+1}^\pm \chi_l^\pm \mp \frac{i}{2} \chi_1^\pm \chi_L^\pm,\label{eq:H_G_pm}
\end{align}
where $L=2N$. For this twisted theory, we use the following half-integer mode expansion with the antiperiodic momentum grid:
\begin{align}
    &\chi_l^\pm = \sqrt{\frac{2}{L}} \sum_{k=1}^{L} e^{\pm i 2\pi \left(k-\frac{1}{2}\right) l/L } d_{k-\frac{1}{2}}^\pm,\label{eq:mode_expansion_chi_NSNS}\\
    &d_{k-\frac{1}{2}}^\pm = \frac{1}{\sqrt{2L}} \sum_{l=1}^{L} e^{\mp i 2\pi \left(k-\frac{1}{2}\right) l / L} \chi_l^\pm,\label{eq:mode_expansion_inverse_NSNS}
\end{align}
The Hermiticity of the lattice Majorana fermions $\chi_l^\pm$ implies
\begin{align}
    d_{k-\frac{1}{2}}^\pm = \left( d_{-k+\frac{1}{2}}^\pm \right)^\dagger.
\end{align}
The Fourier modes satisfy the anticommutation relation,
\begin{align}
    \left\{
    d_{k-\frac{1}{2}}^s, d_{-k'+\frac{1}{2}}^{s'}
    \right\}
    = \delta_{k,k'} \delta_{s,s'}.
\end{align}

Substituting the mode expansion into Eq. (\ref{eq:H_G_pm}), we obtain
\begin{align}
     H_\mathsf{G}^\pm = 2\sum_{k=1}^N \sin \left(
     \frac{2\pi}{L} \left(k-\frac{1}{2}\right)
     \right)
     \left( d_{k-\frac{1}{2}}^\pm \right)^\dagger
     d_{k-\frac{1}{2}}^\pm
     + \text{const.}
\end{align}
The two low-energy regions lie near the continuum points at momentum $k=1$ and momentum $k=N$. Keeping modes with $1\ll\Lambda\ll L$, we find
\begin{align}
    H_\mathsf{G}^\pm \simeq \frac{4\pi}{L} \sum_{k=1}^\Lambda \left(k-\frac{1}{2}\right) \left(d_{k-\frac{1}{2}}^\pm\right)^\dagger d_{k-\frac{1}{2}}^\pm
    + \frac{4\pi}{L} \sum_{k=1}^\Lambda \left(k-\frac{1}{2}\right) \left(d_{N-k+\frac{1}{2}}^\pm\right)^\dagger d_{N-k+\frac{1}{2}}^\pm
    + \text{const.}
\end{align}
Comparing this expression with the continuum Hamiltonians in Eqs. (\ref{eq:H_Maj_plus}) and (\ref{eq:H_Maj_minus}), with $\nuL=\nuR=\frac{1}{2}$, we identify
\begin{align}
    &\chiRkhalf^\pm = d_{k-\frac{1}{2}}^\pm,\label{eq:identification_NSNS_1}\\
    &\chiLkhalf^\pm = d_{N-k+\frac{1}{2}}^\pm.\label{eq:identification_NSNS_2}
\end{align}

We next apply the fermion parity symmetries in Eqs. (\ref{eq:strong_fermion_parity_plus_even}) and (\ref{eq:strong_fermion_parity_minus_even}) to the Fourier modes. Their actions are
\begin{align}
    &(-1)^{F^+} d_{k-\frac{1}{2}}^\pm (-1)^{F^+} = \mp d_{k-\frac{1}{2}}^\pm,\\
    &(-1)^{F^-} d_{k-\frac{1}{2}}^\pm (-1)^{F^-} = \pm d_{k-\frac{1}{2}}^\pm,\\
    &(-1)^{F^\text{w}} d_{k-\frac{1}{2}}^\pm (-1)^{F^\text{w}} = -d_{k-\frac{1}{2}}^\pm,
\end{align}
The NSNS-sector translation symmetries in Eqs. (\ref{eq:TNSNS_plus})-(\ref{eq:TNSNS_w}) act as
\begin{align}
    &T_{\text{NSNS}}^+ d_{k-\frac{1}{2}}^+ \left(T_{\text{NSNS}}^+\right)^{-1} = e^{i\pi \left(k-\frac{1}{2}\right)/N} d_{k-\frac{1}{2}}^+, \quad
    T_{\text{NSNS}}^+ d_{k-\frac{1}{2}}^- \left(T_{\text{NSNS}}^+\right)^{-1} = d_{k-\frac{1}{2}}^-,\\
    &T_{\text{NSNS}}^- d_{k-\frac{1}{2}}^+ \left(T_{\text{NSNS}}^-\right)^{-1} = d_{k-\frac{1}{2}}^+, \quad
    T_{\text{NSNS}}^- d_{k-\frac{1}{2}}^- \left(T_{\text{NSNS}}^-\right)^{-1} = e^{-i\pi \left(k-\frac{1}{2}\right)/N} d_{k-\frac{1}{2}}^-,\\
    &T_{\text{NSNS}}^\text{w} d_{k-\frac{1}{2}}^+ \left(T_{\text{NSNS}}^\text{w}\right)^{-1} = e^{i\pi \left(k-\frac{1}{2}\right)/N} d_{k-\frac{1}{2}}^+, \quad
    T_{\text{NSNS}}^\text{w} d_{k-\frac{1}{2}}^- \left(T_{\text{NSNS}}^\text{w}\right)^{-1} = e^{-i\pi \left(k-\frac{1}{2}\right)/N} d_{k-\frac{1}{2}}^-,
\end{align}
Using the identifications, Eqs. (\ref{eq:identification_NSNS_1}) and (\ref{eq:identification_NSNS_2}), we obtain
\begin{align}
    T_{\text{NSNS}}^+:&\quad 
    \chi_{\text{R},k-\frac{1}{2}}^+ \rightarrow e^{i2\pi \left(k-\frac{1}{2}\right)/L} \chi_{\text{R},k-\frac{1}{2}}^+, \quad
    \chi_{\text{L},k-\frac{1}{2}}^+ \rightarrow - e^{-i2\pi \left(k-\frac{1}{2}\right)/L} \chi_{\text{L},k-\frac{1}{2}}^+,\\
    & \quad \chi_{\text{R},k-\frac{1}{2}}^- \rightarrow \chi_{\text{R},k-\frac{1}{2}}^-, \quad
    \chi_{\text{L},k-\frac{1}{2}}^- \rightarrow \chi_{\text{L},k-\frac{1}{2}}^-,\\
    T_{\text{NSNS}}^-:&\quad
    \chi_{\text{R},k-\frac{1}{2}}^+ \rightarrow \chi_{\text{R},k-\frac{1}{2}}^+,\quad
    \chi_{\text{L},k-\frac{1}{2}}^+ \rightarrow \chi_{\text{L},k-\frac{1}{2}}^+,\\
    &\quad
    \chi_{\text{R},k-\frac{1}{2}}^- \rightarrow e^{-i2\pi \left(k-\frac{1}{2}\right)/L} \chi_{\text{R},k-\frac{1}{2}}^-, \quad
    \chi_{\text{L},k-\frac{1}{2}}^- \rightarrow -e^{i2\pi \left(k-\frac{1}{2}\right)/L} \chi_{\text{L},k-\frac{1}{2}}^-,\\
    T_{\text{NSNS}}^\text{w}:&\quad
    \chi_{\text{R},k-\frac{1}{2}}^+ \rightarrow e^{i2\pi \left(k-\frac{1}{2}\right)/L} \chi_{\text{R},k-\frac{1}{2}}^+, \quad
    \chi_{\text{L},k-\frac{1}{2}}^+ \rightarrow - e^{-i2\pi \left(k-\frac{1}{2}\right)/L} \chi_{\text{L},k-\frac{1}{2}}^+,\\
    &\quad \chi_{\text{R},k-\frac{1}{2}}^- \rightarrow e^{-i2\pi \left(k-\frac{1}{2}\right)/L} \chi_{\text{R},k-\frac{1}{2}}^-, \quad
    \chi_{\text{L},k-\frac{1}{2}}^- \rightarrow -e^{i2\pi \left(k-\frac{1}{2}\right)/L} \chi_{\text{L},k-\frac{1}{2}}^-.
\end{align}
These transformation rules determine the operator structure of the continuum dictionaries. With the phase convention fixed by the exact NSNS lattice algebra, they take the form
\begin{align}
    &T_\text{NSNS}^+ = (-1)^{F_\text{L}^+} e^{i2\pi P^+/L},\label{eq:relation_TNSNSp_fermion_parity_L_plus_momentum}\\
    &T_\text{NSNS}^- = (-1)^{F_\text{L}^-} e^{-i2\pi P^-/L},\label{eq:relation_TNSNSm_fermion_parity_L_minus_momentum}\\
    &T_\text{NSNS}^\text{w} = (-1)^{F_\text{L}^+} (-1)^{F_\text{L}^-} e^{i2\pi P^+/L} e^{-i2\pi P^-/L}.\label{eq:relation_TNSNSw_fermion_parity_L_pm_momentum}
\end{align}
As in the RR sector, a microscopic lattice translation is therefore represented in the Hamiltonian IR limit by a continuum translation dressed by a left-moving fermion parity. In the NSNS sector, however, the absence of zero modes implies that the chiral parities are fermion-even operators. They can therefore be realized without the zero-mode subtleties encountered in the RR sector.

Using these dictionaries, together with the NSNS-sector algebra in Eqs. (\ref{eq:algebra_NSNS_start})-(\ref{eq:algebra_NSNS_end}), we obtain
\begin{align}
    &e^{\pm i 2\pi P^\pm} = (-1)^{F^\pm},\\
    &e^{i 2\pi (P^+ - P^-)} = (-1)^{F^\text{w}},\\
    &(-1)^{F^+} (-1)^{F_\text{L}^+} = (-1)^{F_\text{L}^+} (-1)^{F^+},\\
    &(-1)^{F^-} (-1)^{F_\text{L}^-} = (-1)^{F_\text{L}^-} (-1)^{F^-},\\
    &(-1)^{F_\text{L}^+} (-1)^{F_\text{L}^-} = (-1)^{F_\text{L}^-} (-1)^{F_\text{L}^+}.
\end{align}
Thus, in contrast to the RR sector, the NSNS symmetry algebra is represented linearly. This does not mean that the NSNS sector is itself nonanomalous. Rather, the contrast between the symmetry realizations in the RR and NSNS spin sectors is the characteristic of the 't Hooft anomaly.

\subsection{NSR Sector}
We now turn to the NSR sector. We consider the same no-defect Hamiltonians as in Eq. (\ref{eq:tmpt_1}), but with an odd number of sites, $L=2N+1$. We use the same mode expansion as in Eqs. (\ref{eq:mode_expansion_chi_RR}) and (\ref{eq:d_k_pm_inverse_RR}), but with $L=2N+1$. 
After normal ordering, the Hamiltonians become
\begin{align}
    H_0^\pm = 2 \sum_{k=1}^{N} \sin \left(
    \frac{2 \pi k}{2N+1}
    \right)
    \left( d_k^\pm \right)^\dagger
    d_k^\pm
    + \text{const.}
\end{align}
Note that $k=0$ is the only self-conjugate momentum. The doubled theory therefore contains one plus-sector and one minus-sector zero modes.

Next we examine the Hamiltonian IR limit at large $L$, expanding at $k=0$, near which the right-moving modes locate, and at $k=N$, near which the left-moving modes sit. Keeping modes below the cutoff $1\ll\Lambda\ll L$, we obtain
\begin{align}
    H_0^\pm \simeq \frac{4\pi}{2N+1} \sum_{k=1}^{\Lambda} k \left(d_k^\pm\right)^\dagger d_k^\pm
    + \frac{4\pi}{2N+1} \sum_{k=1}^\Lambda \left( k-\frac{1}{2} \right) \left(d_{N-k+1}^\pm\right)^\dagger d_{N-k+1}^\pm
    + \text{const.}
\end{align}
Comparing to the continuum Hamiltonians in Eqs. (\ref{eq:H_Maj_plus}) and (\ref{eq:H_Maj_minus}), with $\nuL=\frac{1}{2}$ and $\nuR=0$, we identify:
\begin{align}
    &\chiRk^\pm = d_k^\pm,\\
    &\chiLkhalf^\pm = d_{N-k+1}^\pm.
\end{align}

We then apply the fermion parity symmetries in Eqs. (\ref{eq:strong_fermion_parity_plus_odd}) and (\ref{eq:strong_fermion_parity_minus_odd}) to $d_k^\pm$, obtaining
\begin{align}
    &(-1)^{F^+} d_k^\pm (-1)^{F^+} = \mp d_k^\pm,\\
    &(-1)^{F^-} d_k^\pm (-1)^{F^-} = \pm d_k^\pm,\\
    &(-1)^{F^\text{w}} d_k^\pm (-1)^{F^\text{w}} = -d_k^\pm.
\end{align}
The NSR-sector translation symmetries in Eqs. (\ref{eq:TNSR_plus})-(\ref{eq:TNSR_w}) act as
\begin{align}
    &T_{\text{NSR}}^+ d_k^+ \left(T_{\text{NSR}}^+\right)^{-1} = e^{i 2\pi k/L} d_k^+, \quad
    T_{\text{NSR}}^+ d_k^- \left(T_{\text{NSR}}^+\right)^{-1} = d_k^-,\\
    &T_{\text{NSR}}^- d_k^+ \left(T_{\text{NSR}}^-\right)^{-1} = d_k^+, \quad
    T_{\text{NSR}}^- d_k^- \left(T_{\text{NSR}}^-\right)^{-1} = e^{-i 2\pi k/L} d_k^-,\\
    &T_{\text{NSR}}^\text{w} d_k^+ \left(T_{\text{NSR}}^\text{w}\right)^{-1} = e^{i 2\pi k/L} d_k^+, \quad
    T_{\text{NSR}}^\text{w} d_k^- \left(T_{\text{NSR}}^\text{w}\right)^{-1} = e^{-i 2\pi k/L} d_k^-,
\end{align}
from which we can read
\begin{align}
    &T_\text{NSR}^+ = (-1)^{F_\text{L}^+} e^{i2\pi P^+/L},\\
    &T_\text{NSR}^- = (-1)^{F_\text{L}^-} e^{-i2\pi P^-/L},\\
    &T_\text{NSR}^\text{w} = (-1)^{F_\text{L}^+} (-1)^{F_\text{L}^-} e^{i2\pi P^+/L} e^{-i2\pi P^-/L},
\end{align}
with the phase convention compatible with the microscopic NSR algebra.

Using the dictionaries and the algebra of the NSR theory, we achieve the Hamiltonian IR limit:
\begin{align}
    &e^{\pm i 2\pi P^\pm} = e^{\pm i 2\pi n / 16} (-1)^{F_\text{L}^\pm},\\
    &(-1)^{F^+} (-1)^{F_\text{L}^+} = (-1)^{F_\text{L}^+} (-1)^{F^+},\\
    &(-1)^{F^+} (-1)^{F_\text{L}^-} = (-1)^{F_\text{L}^-} (-1)^{F^+},\\
    &(-1)^{F_\text{L}^+} (-1)^{F_\text{L}^-} = (-1)^{F_\text{L}^-} (-1)^{F_\text{L}^+}.
\end{align}
The mod-$8$ anomalous information encoded in the fractional phase is preserved in the continuum limit.

\subsection{RNS Sector}
Finally, we consider the defect sector with an odd number of sites $L=2N+1$. We take the same defect Hamiltonian $H_\mathsf{G}$ as in Eq. (\ref{eq:H_G_pm}) and apply the half-integer Fourier expansion in Eqs. (\ref{eq:mode_expansion_chi_NSNS}) and (\ref{eq:mode_expansion_inverse_NSNS}), with the system size replaced by $L=2N+1$. The resulting Hamiltonians take the form,
\begin{align}
    H_\mathsf{G}^\pm =
    2\sum_{k=1,2,\cdots,N} \sin \left(
    \frac{2\pi}{L} \left(k-\frac{1}{2}\right)
    \right)
    \left( d_{k-\frac{1}{2}}^\pm \right)^\dagger
    d_{k-\frac{1}{2}}^\pm + \text{const.}
\end{align}
Note that each copy of $H_\mathsf{G}^\pm$ has a single zero mode corresponding to $k=N+1$. This zero mode does not appear in the oscillator sum. We now take the Hamiltonian IR limit with large $L$, expanding $H_\mathsf{G}^\pm$ around $k=1$ and $k=N+1$. Keeping modes below the cutoff satisfying $1\ll \Lambda \ll L$, we obtain
\begin{align}
    H_\mathsf{G}^\pm \simeq
    \frac{4\pi}{L} \sum_{k=1}^\Lambda \left(k-\frac{1}{2}\right) \left(d_{k-\frac{1}{2}}^\pm\right)^\dagger d_{k-\frac{1}{2}}^\pm
    + \frac{4\pi}{L} \sum_{k=1}^\Lambda k \left( d_{N+\frac{1}{2}-k}^\pm \right)^\dagger d_{N+\frac{1}{2}-k}^\pm
    + \text{const.}
\end{align}
Comparing to the continuum Hamiltonians in Eqs. (\ref{eq:H_Maj_plus}) and (\ref{eq:H_Maj_minus}), we identify
\begin{align}
    &\chiLk^\pm = d_{N+\frac{1}{2}-k}^\pm,\\
    &\chiRkhalf^\pm = d_{k-\frac{1}{2}}^\pm,
\end{align}
with $\nuL = 0$ and $\nuR = \frac{1}{2}$, as an appropriate RNS theory.

We then act fermion parity symmetries in Eqs. (\ref{eq:strong_fermion_parity_plus_odd}) and (\ref{eq:strong_fermion_parity_minus_odd}), and translation symmetries in Eqs. (\ref{eq:TRNS_plus})-(\ref{eq:TRNS_w}) to the Fourier modes, resulting in
\begin{align}
    &(-1)^{F^+} d_{k-\frac{1}{2}}^\pm (-1)^{F^+} = \mp d_{k-\frac{1}{2}}^\pm,\\
    &(-1)^{F^-} d_{k-\frac{1}{2}}^\pm (-1)^{F^-} = \pm d_{k-\frac{1}{2}}^\pm,\\
    &(-1)^{F^\text{w}} d_{k-\frac{1}{2}}^\pm (-1)^{F^\text{w}} = -d_{k-\frac{1}{2}}^\pm,
\end{align}
and
\begin{align}
    &T_{\text{RNS}}^+ d_{k-\frac{1}{2}}^+ \left(T_{\text{RNS}}^+\right)^{-1} = e^{i 2\pi \left(k-\frac{1}{2}\right)/L} d_{k-\frac{1}{2}}^+, \quad
    T_{\text{RNS}}^+ d_{k-\frac{1}{2}}^- \left(T_{\text{RNS}}^+\right)^{-1} = d_{k-\frac{1}{2}}^-,\\
    &T_{\text{RNS}}^- d_{k-\frac{1}{2}}^+ \left(T_{\text{RNS}}^-\right)^{-1} = d_{k-\frac{1}{2}}^+, \quad
    T_{\text{RNS}}^- d_{k-\frac{1}{2}}^- \left(T_{\text{RNS}}^-\right)^{-1} = e^{-i 2\pi \left(k-\frac{1}{2}\right)/L} d_{k-\frac{1}{2}}^-,\\
    &T_{\text{RNS}}^\text{w} d_{k-\frac{1}{2}}^+ \left(T_{\text{RNS}}^\text{w}\right)^{-1} = e^{i 2\pi \left(k-\frac{1}{2}\right)/L} d_{k-\frac{1}{2}}^+, \quad
    T_{\text{RNS}}^\text{w} d_{k-\frac{1}{2}}^- \left(T_{\text{RNS}}^\text{w}\right)^{-1} = e^{-i 2\pi \left(k-\frac{1}{2}\right)/L} d_{k-\frac{1}{2}}^-.
\end{align}
These transformation rules determine the operator structure of the continuum dictionaries:
\begin{align}
    &T_\text{RNS}^+ = (-1)^{F_\text{L}^+} e^{i2\pi P^+/L},\label{eq:relation_TRNSp_fermion_parity_L_plus_momentum}\\
    &T_\text{RNS}^- = (-1)^{F_\text{L}^-} e^{-i2\pi P^-/L},\label{eq:relation_TRNSm_fermion_parity_L_minus_momentum}\\
    &T_\text{RNS}^\text{w} = -i(-1)^{F_\text{L}^+} (-1)^{F_\text{L}^-} e^{i2\pi P^+/L} e^{-i2\pi P^-/L},\label{eq:relation_TRNSw_fermion_parity_L_pm_momentum}
\end{align}
Using these dictionaries, together with the microscopic RNS algebra in Eqs. (\ref{eq:algebra_RNS_start})-(\ref{eq:algebra_RNS_end}), we obtain the algebraic structure in the Hamiltonian IR limit,
\begin{align}
    & e^{\pm i 2\pi P^\pm} = e^{\pm i 2\pi n' / 16} (-1)^{F_\text{R}^\pm},\\
    & (-1)^{F^+} (-1)^{F_\text{L}^+} = (-1)^{F_\text{L}^+} (-1)^{F^+},\\
    & (-1)^{F^+} (-1)^{F_\text{L}^-} = - (-1)^{F_\text{L}^-} (-1)^{F^+},\\
    & (-1)^{F_\text{L}^+} (-1)^{F_\text{L}^-} = - (-1)^{F_\text{L}^-} (-1)^{F_\text{L}^+}.
\end{align}
We remark that the RNS sector realizes the opposite mod-$8$ anomaly class from the NSR sector. In addition, comparing with the corresponding NSR relations , we find that the RNS symmetry algebra contains nontrivial minus signs that are absent in the NSR sector. These signs indicate a projective realization of the symmetry algebra and constitute a manifestation of the ’t Hooft anomaly. This anomaly is intrinsic to the dissipative Majorana chain. Indeed, for a closed Majorana chain with an odd number of Majorana fermions, fermion parity is an outer automorphism and cannot be represented by an operator within a single irreducible Hilbert space. In the doubled Liouville space, however, the ket and bra sectors together allow the strong fermion-parity transformations to be represented as operators, and their algebra exhibits the anomalous minus signs.

\section{Bosonization}\label{sec:bosonization}
Recent studies have shown that, in closed systems, an invertible anomalous symmetry can lead to a non-invertible symmetry after lattice bosonization or gauging\cite{10.21468/SciPostPhys.16.3.064,Seiberg:2024gek}. Here we ask whether the same anomalous structure also appears in the open quantum systems. Specifically we focus on our Lindbladian Majorana system in which the strong translation symmetries obey anomalous algebraic relations with the strong fermion parity symmetries; in particular they anticommute as in Eq. (\ref{eq:strong_parity_and_T_anticommute}). Sections \ref{sec:JW_and_spin_op_in_ket_bra_sectors}-\ref{sec:non_invertible_symm} concerns even system size $L=2N$; Sec. \ref{sec:bosonization_odd} treats odd system size separately.

In Sec. \ref{sec:JW_and_spin_op_in_ket_bra_sectors}, we derive the relations between Majorana fermions and spin operators using Jordan--Wigner transformation, and then use modular conjugation to construct the corresponding spin operators in the ket and bra subspaces. We then, in Sec. \ref{sec:bosonization_based_on_fermion_parity_symm}, apply the bosonization known as the Gliozzi--Scherk--Olive (GSO) projection to the Majorana Lindbladians, focusing on the Ising subspace which yields an Ising type Lindbladian. Finally, in Sec. \ref{sec:non_invertible_symm}, we build up non-invertible symmetries $\widetilde{\mathsf{D}}^{\pm}$ for the Ising Lindbladian, based on the anomalous relations between strong translation symmetries and strong fermion parity symmetries.

\subsection{Jordan--Wigner Transformation and Spin Operators in Ket/Bra Sectors}\label{sec:JW_and_spin_op_in_ket_bra_sectors}
For even $L=2N$, we introduce a spin representation of the Lindbladian Majorana system by applying the following Jordan--Wigner transformation, mapping $2L=4N$ Majorana operators to Pauli operators on $2N$ spin sites,
\begin{align}
    &\chi_{2j-1}^{+} = - \left( \prod_{k=1}^{j-1} Z_{k} \right) X_{j},\label{eq:JW_1}\\
    &\chi_{2j}^{+} = \left( \prod_{k=1}^{j-1} Z_{k} \right) Y_{j},\\
    &\chi_{2j-1}^{-} = \left(\prod_{k=1}^{N} Z_{k}\right)  \left( \prod_{k=1}^{j-1} Z_{N+k} \right) X_{N+j},\\
    &\chi_{2j}^{-} = \left(\prod_{k=1}^{N} Z_{k}\right) \left( \prod_{k=1}^{j-1} Z_{N+k} \right) Y_{N+j},\label{eq:JW_4}
\end{align}
where $j=1,2,\cdots,N$. The inverse transformation is derived straightforwardly from Eqs. (\ref{eq:JW_1})-(\ref{eq:JW_4}),
\begin{align}
    &X_{j} = - \left( \prod_{k=1}^{j-1} \left( i \chi_{2k-1}^{+} \chi_{2k}^{+} \right) \right) \chi_{2j-1}^{+},\\
    &Y_{j} = \left( \prod_{k=1}^{j-1} \left( i \chi_{2k-1}^{+} \chi_{2k}^{+} \right) \right) \chi_{2j}^{+},\\
    &X_{N+j} = (-1)^{F^{+}} \left( \prod_{k=1}^{j-1} \left( -i \chi_{2k-1}^{-} \chi_{2k}^{-} \right) \right) \chi_{2j-1}^{-},\\
    &Y_{N+j} = (-1)^{F^{+}} \left( \prod_{k=1}^{j-1} \left( -i \chi_{2k-1}^{-} \chi_{2k}^{-} \right) \right) \chi_{2j}^{-},
\end{align}
again with $j=1,2,\cdots,N$.

In this spin representation, the strong and weak fermion parity symmetries are,
\begin{align}
    &(-1)^{F^{+}} = \prod_{l=1}^{N} \left( i \chi_{2l-1}^{+} \chi_{2l}^{+} \right) = \prod_{j=1}^{N} Z_{j},\\
    &(-1)^{F^{-}} = \mathcal{J} (-1)^{F^{+}} \mathcal{J}^{-1} = \prod_{l=1}^{N} \left( -i \chi_{2l-1}^{-} \chi_{2l}^{-} \right) = \prod_{j=1}^{N} Z_{N+j},\\
    &(-1)^{F^{\text{w}}} = (-1)^{F^{+}} (-1)^{F^{-}} = \prod_{j=1}^{2N}Z_j.
\end{align}

We will use the following identities for Majorana bilinears. For bulk terms, one finds
\begin{align}
    &\chi_{2j-1}^{+} \chi_{2j}^{+} = -i Z_{j},\label{eq:JW_chi_X_Z_1}\\
    &\chi_{2j-1}^{-} \chi_{2j}^{-} = i Z_{N+j},
\end{align}
with $j=1,2,\cdots,N$, and
\begin{align}
    &\chi_{2j}^{+} \chi_{2j+1}^{+} = -i X_{j} X_{j+1},\\
    &\chi_{2j}^{-} \chi_{2j+1}^{-} = i X_{N+j} X_{N+j+1},
\end{align}
with $j=1,2,\cdots,N-1$. For boundary terms, we have the relations
\begin{align}
    &\chi_{2N}^{+} \chi_{1}^{+} = i (-1)^{F^{+}} X_{N} X_{1},\\
    &\chi_{2N}^{-} \chi_{1}^{-} = -i (-1)^{F^{-}} X_{2N} X_{N+1}\label{eq:JW_chi_X_Z_end}.
\end{align}

We next reorganize the $2N$ spin sites into ket and bra spin degrees of freedom. The first $N$ spin sites are identified with the ket sector, while the remaining $N$ spin sites are related to the bra sector by modular conjugation:
\begin{align}
    &Z_{j}^{+} \equiv Z_{j},\\
    &Z_{j}^{-} \equiv \mathcal{J} Z_{j}^{+} \mathcal{J}^{-1} = \mathcal{J} \left( i \chi_{2j-1}^{+} \chi_{2j}^{+} \right) \mathcal{J}^{-1} = -i \chi_{2j-1}^{-} \chi_{2j}^{-} = Z_{N+j},\\
    &X_{j}^{+} \equiv V X_{j},\\
    &X_{j}^{-} \equiv \mathcal{J} X_{j}^{+} \mathcal{J}^{-1} = - (-1)^{F^{+}} V^\dagger X_{N+j},\\
    &Y_{j}^{+} \equiv V Y_{j},\\
    &Y_{j}^{-} \equiv \mathcal{J} Y_j^+ \mathcal{J}^{-1} =  (-1)^{F^{+}} V^\dagger Y_{N+j},
\end{align}
where the unitary dressing $V\equiv \left( 1+i (-1)^{F^\text{w}} \right)/\sqrt{2}$ ensures that every spin operator in the ket space commutes with every spin operator in the bra space.  With these definitions, note that both $Z_{j}^{+}$ and $Z_{j}^{-}$ are even in strong fermion parities, meaning that $[Z_{j}^{+},(-1)^{F^{\pm}}]=0$ and $[Z_{j}^{-},(-1)^{F^{\pm}}]=0$. By contrast,
\begin{align}
    &\{X_{j}^{\pm},(-1)^{F^{\pm}}\} = \{Y_{j}^{\pm},(-1)^{F^{\pm}}\} = 0,\\
    &[X_{j}^{\pm},(-1)^{F^{\mp}}] = [Y_{j}^{\pm},(-1)^{F^{\mp}}] = 0.
\end{align}

\subsection{Bosonization Based on Fermion Parity Symmetry (Even $L=2N$)}\label{sec:bosonization_based_on_fermion_parity_symm}
Bosonization on a lattice can be implemented in two steps: first summing over spin structures, and then imposing a projection\cite{10.21468/SciPostPhys.16.3.064,karch2019web}. In the present setting, the two spin structures correspond to the $\mathsf{G}$-defect and no-defect sector, which in this section is denoted with a subscript $0$. We therefore begin with the direct-sum Hilbert space $\widetilde{\mathcal{H}} = \mathcal{H}_{\mathsf{G}} \oplus \mathcal{H}_0$, whose dimension is $2^{N+1}$. For a closed system, the corresponding Hamiltonian is naturally defined as: $\widetilde{H} = H_{\mathsf{G}} \oplus H_0 \in \mathcal{B}(\widetilde{\mathcal{H}})$, a $2^{N+1}\times 2^{N+1}$ linear map. For an open quantum system, we extend the same idea to the jump operators $\widetilde{L} = L_{\mathsf{G}} \oplus L_0 \in \mathcal{B}(\widetilde{\mathcal{H}})$. Here for notational simplicity, we suppress the jump operator subscript $m$. Given $\widetilde{H}$ and $\widetilde{L}$, the corresponding Lindbladian $\widetilde{\mathcal{L}}$ is constructed in the standard form (Eq. (\ref{eq:Lindblad_eq})).

In the double Hilbert space formalism, a density matrix $\rho \in \mathcal{B}(\widetilde{\mathcal{H}})$ is regarded as a vector in the Liouville space
\begin{align}
    \widetilde{\mathcal{K}} &= (\mathcal{H}_{\mathsf{G}} \oplus \mathcal{H}_0) \otimes (\mathcal{H}_{\mathsf{G}} \oplus \mathcal{H}_0)\\
    &= \mathcal{H}_{\mathsf{G}} \otimes \mathcal{H}_{\mathsf{G}}
    \oplus \mathcal{H}_{\mathsf{G}} \otimes \mathcal{H}_{0}
    \oplus \mathcal{H}_{\mathsf{0}} \otimes \mathcal{H}_{G}
    \oplus \mathcal{H}_{\mathsf{0}} \otimes \mathcal{H}_{0}
\end{align}
which has $\text{dim}(\widetilde{\mathcal{K}})=2^{2N+2}$. This space naturally decomposes into four sectors. Accordingly, operators inherited from the direct-sum constructions, such as $\widetilde{\mathcal{L}}$, $\widetilde{H}^{\pm}$ and $\widetilde{L}^{\pm}$, take a $4\times 4$ block-diagonal form. For instance,
\begin{align}
    \widetilde{H}^{+} &= \widetilde{H} \otimes \widetilde{I} = (H_{\mathsf{G}} \oplus H_0) \otimes (I_{\mathsf{G}} \oplus I_0)\\
    &= H_{\mathsf{G}} \otimes I_{\mathsf{G}} \oplus H_{\mathsf{G}} \otimes I_0 \oplus H_0 \otimes I_{\mathsf{G}} \oplus H_0 \otimes I_0\\
    &= H_{\mathsf{G}}^+ \oplus H_{\mathsf{G}}^+ \oplus H_0^+ \oplus H_0^+,\label{eq:tildeH_plus_direct_sum_four_subspaces}
\end{align}
and similarly,
\begin{align}
    \widetilde{H}^- &= I_{\mathsf{G}}\otimes H_{\mathsf{G}} \oplus I_{\mathsf{G}}\otimes H_0 \oplus I_0\otimes H_{\mathsf{G}} \oplus I_0\otimes H_0 \label{eq:tildeH_minus_direct_sum_four_subspaces}\\
    &= H_{\mathsf{G}}^- \oplus H_0^- \oplus H_{\mathsf{G}}^- \oplus H_0^-,\\
    \widetilde{L}^+ \widetilde{L}^- &= L_{\mathsf{G}}^+ L_{\mathsf{G}}^- \oplus L_{\mathsf{G}}^+ L_0^- \oplus L_0^+ L_{\mathsf{G}}^- \oplus L_0^+ L_0^-,\label{eq:tilde_LpLm}\\
    \widetilde{L}^{+\dagger}\widetilde{L}^+ &= L_{\mathsf{G}}^{+\dagger} L_{\mathsf{G}}^+ \oplus L_{\mathsf{G}}^{+\dagger} L_{\mathsf{G}}^+ \oplus L_0^{+\dagger} L_0^+ \oplus L_0^{+\dagger} L_0^+,\\
    \widetilde{L}^{-\dagger} \widetilde{L}^- &= L_{\mathsf{G}}^{-\dagger} L_{\mathsf{G}}^- \oplus L_0^{-\dagger} L_0^- \oplus L_{\mathsf{G}}^{-\dagger} L_{\mathsf{G}}^- \oplus L_0^{-\dagger} L_0^-.\label{eq:tilde_LmLm}
\end{align}
In what follows, we take $\mathsf{G}$ to be the defect associated with the weak fermion parity symmetry to construct all the building blocks $H_\mathsf{G}^\pm$ and $L_\mathsf{G}^\pm$. That being said, the subscripts $\mathsf{G}$ and $0$ correspond to NSNS and RR, respectively. Equations (\ref{eq:tilde_LpLm})-(\ref{eq:tilde_LmLm}) are valid for both fermion-parity even and odd jump operators. In the later case, this can be seen by using the representations $\widetilde{L}^+=\widetilde{L}\otimes\widetilde{I}$ and $\widetilde{L}^- = \left[(-1)^F\widetilde{I}\right] \otimes \widetilde{L}$. However, fermion-parity-odd operators generally give rise to highly nonlocal string operators under the Jordan--Wigner transformation, so that the resulting bosonization does not take a neat form. Therefore, we focus on the fermion-parity even jump terms in what follows. Combining the Hamiltonian and jump operator contributions above, the full Lindbladian $\widetilde{\mathcal{L}}$ is
\begin{align}
    \widetilde{\mathcal{L}} = -i \left(\widetilde{H}^+ - \widetilde{H}^-\right) + \widetilde{L}^+ \widetilde{L}^- -\frac{1}{2} \widetilde{L}^{+\dagger}\widetilde{L}^+ - \frac{1}{2}\widetilde{L}^{-\dagger}\widetilde{L}^-.\label{eq:Lindbladian_aux_space}
\end{align}

Our goal is to extract an Ising type Lindbladian from $\widetilde{\mathcal{L}}$. As a concrete example, we consider the energy dephasing jump operator in Eq. (\ref{eq:L0_collective_density_dephasing_jump_op}). We first rewrite the Hamiltonians $H_{\mathsf{G}/0}^{\pm}$ and the jump operator $L_{\mathsf{G}/0}^{\pm}$ in terms of spin variables using the Jordan--Wigner relations derived in Sec. \ref{sec:JW_and_spin_op_in_ket_bra_sectors}. For the no-defect sector, one obtains
\begin{align}
    H_0^{+} =& -\frac{i}{2} \sum_{l=1}^{L-1} \chi_{l}^{+} \chi_{l+1}^{+} - \frac{i}{2} \chi_{L}^{+} \chi_{1}^{+}\\
    =& -\frac{i}{2} \sum_{j=1}^{N} \left(-i Z_{j}^{+}\right) - \frac{i}{2} \sum_{j=1}^{N-1} \left(-i X_{j}^{+}X_{j+1}^{+}\right) - \frac{i}{2} i (-1)^{F^{+}} X_{N}^{+}X_{1}^{+}\\
    =& -\frac{1}{2} \sum_{j=1}^{N} Z_{j}^{+} - \frac{1}{2} \sum_{j=1}^{N-1} X_{j}^{+}X_{j+1}^{+} + \frac{1}{2} (-1)^{F^{+}} X_{N}^{+}X_{1}^{+},\\
    H_0^{-} =& \frac{i}{2} \sum_{l=1}^{L-1} \chi_{l}^{-} \chi_{l+1}^{-} + \frac{i}{2} \chi_{L}^{-} \chi_{1}^{-}\\
    =& -\frac{1}{2} \sum_{j=1}^{N} Z_{j}^{-} - \frac{1}{2} \sum_{j=1}^{N-1} X_{j}^{-}X_{j+1}^{-} + \frac{1}{2} (-1)^{F^{-}} X_{N}^{-}X_{1}^{-}.
\end{align}
Similarly, for the defect $H_{\mathsf{G}}^{\pm}$, no-defect $L_0^{\pm}$ and defect $L_{\mathsf{G}}^{\pm}$, one obtain
\begin{align}
    H_{\mathsf{G}}^{\pm} &= -\frac{1}{2} \sum_{j=1}^{N} Z_{j}^{\pm} - \frac{1}{2} \sum_{j=1}^{N-1} X_{j}^{\pm}X_{j+1}^{\pm} - \frac{1}{2} (-1)^{F^{\pm}} X_{N}^{\pm}X_{1}^{\pm},\\
    L_{0}^{\pm} &= \sqrt{\gamma}\left(
    -\sum_{j=1}^{N} Z_j^{\pm} - \sum_{j=1}^{N-1} X_j^{\pm}X_{j+1}^{\pm} + (-1)^{F^{\pm}} X_{N}^{\pm} X_1^{\pm}
    \right),\\
    L_{\mathsf{G}}^{\pm} &= \sqrt{\gamma}\left(
    -\sum_{j=1}^{N} Z_j^{\pm} - \sum_{j=1}^{N-1} X_j^{\pm}X_{j+1}^{\pm} - (-1)^{F^{\pm}} X_{N}^{\pm} X_1^{\pm}
    \right).
\end{align}
To recover the Ising sector, we impose the GSO projection---we require $(-1)^{F^{+}}=(-1)^{F^{-}}=-1$ in the no-defect sector, whereas $(-1)^{F^{+}}=(-1)^{F^{-}}=1$ in the defect sector. These sector-dependent constraints can be implemented by introducing
\begin{align}
    \widetilde{\eta} &= \begin{pmatrix}
        1 & 0\\
        0 & -1
    \end{pmatrix}_{\mathsf{G}/0},\\
    \widetilde{\eta}^+ &= \widetilde{\eta} \otimes\widetilde{I},\\
    \widetilde{\eta}^- &= \widetilde{I} \otimes \widetilde{\eta},\\
    \widetilde{(-1)^{F^+}} &= \text{diag}\left((-1)^{F^+},(-1)^{F^+},(-1)^{F^+},(-1)^{F^+}\right)\\
    \widetilde{(-1)^{F^-}} &= \text{diag}\left((-1)^{F^-},(-1)^{F^-},(-1)^{F^-},(-1)^{F^-}\right).
\end{align}
The two projection conditions are then $\widetilde{\eta}^+ \widetilde{(-1)^{F^+}}=1$ and $\widetilde{\eta}^- \widetilde{(-1)^{F^-}}=1$. We remark that each condition reduces the Liouville space by a factor of two, so that the projected Liouville space has dimension $2^{2N}$, as expected for the Liouville space of a $N$-spin open quantum system.

We now seek an equivalent Lindbladian $\widetilde{\mathcal{L}_{\text{Ising}}}$ with the following properties. First, $\widetilde{\mathcal{L}_{\text{Ising}}}$ should be built from local operators that commute with the projection $P_{\text{Ising}}=\frac{1+\widetilde{\eta}^+ \widetilde{(-1)^{F^+}}}{2} \frac{1+\widetilde{\eta}^- \widetilde{(-1)^{F^-}}}{2}$, so that their action preserves the projected subspace. Second, after imposing the projection, $\tilde{\mathcal{L}}$ should take the form of an Ising open quantum system. To achieve these, we define the local operators
\begin{align}
    \widetilde{Z}_j^+ &= \text{diag}\left( Z_j^+, Z_j^+, Z_j^+, Z_j^+ \right),\\
    \widetilde{X}_j^+ &= \begin{pmatrix}
        0 & 0 & X_j^+ & 0\\
        0 & 0 & 0 & X_j^+\\
        X_j^+ & 0 & 0 & 0\\
        0 & X_j^+ & 0 & 0
    \end{pmatrix},\\
    \widetilde{Z}_j^- &= \text{diag}\left( Z_j^-, Z_j^-, Z_j^-, Z_j^- \right),\\
    \widetilde{X}_j^- &= \begin{pmatrix}
        0 & X_j^- & 0 & 0\\
        X_j^- & 0 & 0 & 0\\
        0 & 0 & 0 & X_j^-\\
        0 & 0 & X_j^- & 0
    \end{pmatrix},
\end{align}
which all commute with the projection $P_{\text{Ising}}$. We also define the modular conjugation on the four-sector Liouville space as,
\begin{align}
    \widetilde{\mathcal{J}} = \begin{pmatrix}
        \mathcal{J} & 0 & 0 & 0\\
        0 & 0 & \mathcal{J} & 0\\
        0 & \mathcal{J} & 0 & 0\\
        0 & 0 & 0 & \mathcal{J}
    \end{pmatrix},\label{eq:tilde_J}
\end{align}
which maps $\widetilde{Z}_j^\pm$ to $\widetilde{Z}_j^\mp$ and $\widetilde{X}_j^\pm$ to $\widetilde{X}_j^\mp$.

In terms of these local operators, (\ref{eq:tildeH_plus_direct_sum_four_subspaces}) and (\ref{eq:tildeH_minus_direct_sum_four_subspaces}) can be rewritten as
\begin{align}
    \widetilde{H}^+ &= -\frac{1}{2} \sum_{j=1}^{N} \widetilde{Z}_j^+ - \frac{1}{2} \sum_{j=1}^{N-1} \widetilde{X}_j^+ \widetilde{X}_{j+1}^+ - \frac{1}{2} \widetilde{\eta}^+ \widetilde{(-1)^{F^+}} \widetilde{X}_N^+ \widetilde{X}_1^+,\\
    \widetilde{H}^- &= -\frac{1}{2} \sum_{j=1}^{N} \widetilde{Z}_j^- - \frac{1}{2} \sum_{j=1}^{N-1} \widetilde{X}_j^- \widetilde{X}_{j+1}^- - \frac{1}{2} \widetilde{\eta}^- \widetilde{(-1)^{F^-}} \widetilde{X}_N^- \widetilde{X}_1^-.
\end{align}
After imposing the projection $P_{\text{Ising}}$, the boundary parity factors are replaced by the corresponding sector labels, and the Hamiltonians take the standard Ising form. The same reasoning applies to the jump operators, giving
\begin{align}
    \widetilde{L}^{\pm} &= \sqrt{\gamma}\left(
    -\sum_{j=1}^{N} \widetilde{Z}_j^{\pm} - \sum_{j=1}^{N-1} \widetilde{X}_j^{\pm} \widetilde{X}_{j+1}^{\pm} - \widetilde{\eta}^{\pm} \widetilde{(-1)^{F^{\pm}}} \widetilde{X}_N^{\pm} \widetilde{X}_1^{\pm}
    \right).
\end{align}
Finally, projecting the full Lindbladian $\widetilde{\mathcal{L}}$ in Eq. (\ref{eq:Lindbladian_aux_space}) results in
\begin{align}
    \widetilde{\mathcal{L}}_\text{Ising} = 
    -i \left(H_{\text{Ising}}^+ - H_{\text{Ising}}^-\right) + L_{\text{Ising}}^+ L_{\text{Ising}}^- -\frac{1}{2} L_{\text{Ising}}^{+\dagger}L_{\text{Ising}}^+ - \frac{1}{2}L_{\text{Ising}}^{-\dagger}L_{\text{Ising}}^-,
\end{align}
where
\begin{align}
     H_{\text{Ising}}^{\pm} &= -\frac{1}{2} \sum_{j=1}^{N} \widetilde{Z}_j^{\pm} - \frac{1}{2} \sum_{j=1}^{N-1} \widetilde{X}_j^{\pm} \widetilde{X}_{j+1}^{\pm} - \frac{1}{2} \widetilde{X}_N^{\pm} \widetilde{X}_1^{\pm},\\
     L_{\text{Ising}}^{\pm} &= \sqrt{\gamma}\left(
     - \sum_{j=1}^{N} \widetilde{Z}_j^{\pm} - \sum_{j=1}^{N-1} \widetilde{X}_j^{\pm} \widetilde{X}_{j+1}^{\pm} - \widetilde{X}_N^{\pm} \widetilde{X}_1^{\pm}
     \right).
\end{align}
Strictly speaking, $H_{\text{Ising}}^{\pm}$ and $L_{\text{Ising}}^{\pm}$ are still written in the enlarged four-sector Liouville space. To obtain the minimal Ising representation, one should further restrict to the image of $P_{\text{Ising}}$, equivalently by applying an isometry that removes the kernel of the projection.

\subsection{Non-Invertible Symmetries}\label{sec:non_invertible_symm}
We now return to the translation symmetries. Motivated by Eqs. (\ref{eq:tildeH_plus_direct_sum_four_subspaces}) and (\ref{eq:tildeH_minus_direct_sum_four_subspaces}), we define strong translation symmetries on the four-sector Liouville space $\widetilde{\mathcal{K}}$ as,
\begin{align}
    \widetilde{T}^+ &= T_\mathsf{G}^+ \oplus T_\mathsf{G}^+ \oplus T_0^+ \oplus T_0^+,\\
    \widetilde{T}^- &= T_\mathsf{G}^- \oplus T_0^- \oplus T_\mathsf{G}^- \oplus T_0^-,
\end{align}
which leave the Hamiltonians $\widetilde{H}^\pm$ invariant under conjugation,
\begin{align}
    \widetilde{T}^+ \widetilde{H}^\pm \left(\widetilde{T}^+\right)^{-1} = \widetilde{H}^\pm,\\
    \widetilde{T}^- \widetilde{H}^\pm \left(\widetilde{T}^-\right)^{-1} = \widetilde{H}^\pm.
\end{align}
Similarly, the jump operator terms $\widetilde{L}^\pm$, and hence the full Lindbladian $\widetilde{\mathcal{L}}$ are invariant under the action of $\widetilde{T}^{\pm}$.

However, both $\widetilde{T}^+$ and $\widetilde{T}^-$ do not commute with $P_{\text{Ising}}$, since $\widetilde{T}^+ \widetilde{(-1)^{F^+}} = \widetilde{\eta}^+ \widetilde{(-1)^{F^+}} \widetilde{T}^+$ and $\widetilde{T}^- \widetilde{(-1)^{F^-}} = \widetilde{\eta}^- \widetilde{(-1)^{F^-}} \widetilde{T}^-$. That said, $\widetilde{T}^+$ and $\widetilde{T}^-$ are not the symmetries of the projected Ising sector.

Nevertheless, we can project out the subspaces that are incompatible with the constraints, i.e. the minus sectors in $\widetilde{\eta}^\pm$. We define
\begin{align}
    \widetilde{\mathsf{D}}^+ &= \frac{1+\widetilde{\eta}^+}{2} \widetilde{T}^+ = \begin{pmatrix}
        T_\mathsf{G}^+ & & & \\
         & T_\mathsf{G}^+ & & \\
         & & 0 & \\
         & & & 0
    \end{pmatrix},\\
    \widetilde{\mathsf{D}}^- &= \frac{1+\widetilde{\eta}^-}{2} \widetilde{T}^- = \begin{pmatrix}
        T_\mathsf{G}^- & & & \\
         & 0 & & \\
         & & T_\mathsf{G}^- & \\
         & & & 0
    \end{pmatrix}.
\end{align}
As we will see, these are the symmetries of $\widetilde{\mathcal{L}}$, but they have nontrivial kernel and therefore are non-invertible symmetries.

Let us first analyze $\widetilde{D}^+$. Using Jordan--Wigner transformation, the diagonal translation operator $T_{\mathsf{G}}^+$, namely $T_{\text{NSNS}}^+$ in Eq. (\ref{eq:TNSNS_plus}), can be written as
\begin{align}
    T_{\mathsf{G}}^+ = e^{-i\frac{N}{4}\pi}
    \left( \frac{1+i Z_1^+}{\sqrt{2}} \right)
    \left( \frac{1+i X_1^+ X_2^+}{\sqrt{2}} \right)
    \left( \frac{1+i Z_2^+}{\sqrt{2}} \right)
    \left( \frac{1+i X_2^+ X_3^+}{\sqrt{2}} \right)
    \cdots
    \left( \frac{1+i Z_N^+}{\sqrt{2}} \right),
\end{align}
which leads to
\begin{align}
    \widetilde{\mathsf{D}}^+ = e^{-i\frac{N}{4}\pi}
    \left( \frac{1+i \widetilde{Z}_1^+}{\sqrt{2}} \right)
    \left( \frac{1+i \widetilde{X}_1^+ \widetilde{X}_2^+}{\sqrt{2}} \right)
    \left( \frac{1+i \widetilde{Z}_2^+}{\sqrt{2}} \right)
    \left( \frac{1+i \widetilde{X}_2^+ \widetilde{X}_3^+}{\sqrt{2}} \right)
    \cdots
    \left( \frac{1+i \widetilde{Z}_N^+}{\sqrt{2}} \right)
    \frac{1+\widetilde{\eta}^+}{2}.
\end{align}

We then act $\widetilde{\mathsf{D}}^+$ on $\widetilde{Z}_j^+$, obtaining
\begin{align}
    \widetilde{\mathsf{D}}^+ \widetilde{Z}_j^+ &= \widetilde{X}_j^+ \widetilde{X}_{j+1}^+ \widetilde{\mathsf{D}}^+, \qquad j=1,2,\cdots,N-1\\
    \widetilde{\mathsf{D}}^+ \widetilde{Z}_N^+ &= \widetilde{X}_N^+ \widetilde{X}_{1}^+ \widetilde{(-1)^{F^+}} \widetilde{\mathsf{D}}^+.
\end{align}
Inside the Ising subspace, the projection condition implies $\widetilde{\eta}^+ \widetilde{(-1)^{F^+}}=1\Rightarrow\widetilde{\eta}^+ = \widetilde{(-1)^{F^+}}$, and so $\widetilde{(-1)^{F^+}} \widetilde{\mathsf{D}}^+ = \widetilde{\eta}^+ \widetilde{\mathsf{D}}^+ = \widetilde{\mathsf{D}}^+$. Thus, after the projection,
\begin{align}
    &\left(
    \widetilde{\mathsf{D}}^+ \widetilde{Z}_j^+ = \widetilde{X}_j^+ \widetilde{X}_{j+1}^+ \widetilde{\mathsf{D}}^+
    \right)\bigg\vert_\text{Ising}, \quad j=1,2,\cdots,N \label{eq:DZ_XXD_Ising}\\
    &\widetilde{X}_{N+1}^+ \sim \widetilde{X}_1^+,
\end{align}
which is precisely the Kramers--Wannier duality. Similarly, acting $\widetilde{\mathsf{D}}^+$ on $\widetilde{X}_j^+ \widetilde{X}_{j+1}^+$ results in
\begin{align}
    &\widetilde{\mathsf{D}}^+ \widetilde{X}_j^+ \widetilde{X}_{j+1}^+ = \widetilde{Z}_{j+1}^+ \widetilde{\mathsf{D}}^+, \qquad j=1,2,\cdots N-1\\
    &\widetilde{\mathsf{D}}^+ \widetilde{X}_N^+ \widetilde{X}_1^+ = \widetilde{Z}_1^+ \widetilde{(-1)^{F^+}} \widetilde{\mathsf{D}}^+.
\end{align}
After imposing the Ising projection, these become
\begin{align}
    &\left(
    \widetilde{\mathsf{D}}^+ \widetilde{X}_j^+ \widetilde{X}_{j+1}^+ = \widetilde{Z}_{j+1}^+ \widetilde{\mathsf{D}}^+
    \right) \bigg\vert_{\text{Ising}}, \qquad j=1,2,\cdots,N \label{eq:DXX_ZD_Ising}\\
    &\widetilde{X}_{N+1}^+ \sim \widetilde{X}_1^+; \quad
    \widetilde{Z}_{N+1}^+ \sim \widetilde{Z}_1^+.
\end{align}
On the other hand, $\widetilde{\mathsf{D}}^+$ acts trivially on local operators in the bra sector:
\begin{align}
    &\widetilde{\mathsf{D}}^+ \widetilde{Z}_j^- = 
    \widetilde{Z}_j^- \widetilde{\mathsf{D}}^+, \label{eq:DZ_ZD}\\
    &\widetilde{\mathsf{D}}^+ \widetilde{X}_j^- \widetilde{X}_{j+1}^- = 
    \widetilde{X}_j^- \widetilde{X}_{j+1}^- \widetilde{\mathsf{D}}^+.\label{eq:DXX_XXD}
\end{align}

Next, to achieve the operation of $\widetilde{\mathsf{D}}^-$ on local operators, we simply apply the modular conjugation $\widetilde{\mathcal{J}}$ defined in Eq. (\ref{eq:tilde_J}) to Eqs. (\ref{eq:DZ_XXD_Ising}), (\ref{eq:DXX_ZD_Ising}), (\ref{eq:DZ_ZD}) and (\ref{eq:DXX_XXD}) (we remark that $\widetilde{\mathcal{J}} \widetilde{\mathsf{D}}^+\widetilde{\mathcal{J}}^{-1} = \widetilde{\mathsf{D}}^-$), obtaining
\begin{align}
    &\left(
    \widetilde{\mathsf{D}}^- \widetilde{Z}_j^- = \widetilde{X}_j^- \widetilde{X}_{j+1}^- \widetilde{\mathsf{D}}^-
    \right) \bigg\vert_{\text{Ising}},\\
    &\left(
    \widetilde{\mathsf{D}}^- \widetilde{X}_j^- \widetilde{X}_{j+1}^- = \widetilde{Z}_{j+1}^- \widetilde{\mathsf{D}}^-
    \right) \bigg\vert_{\text{Ising}},\\
    &\widetilde{\mathsf{D}}^- \widetilde{Z}_j^+ = \widetilde{Z}_j^+ \widetilde{\mathsf{D}}^-,\\
    &\widetilde{\mathsf{D}}^- \widetilde{X}_j^+ \widetilde{X}_{j+1}^+ = \widetilde{X}_j^+ \widetilde{X}_{j+1}^+ \widetilde{\mathsf{D}}^-,
\end{align}
where $j=1,2,\cdots,N$ and $N+1\sim1$. As a result,
\begin{align}
    \left(
    \widetilde{\mathsf{D}}^{\pm} \widetilde{\mathcal{L}} = \widetilde{\mathcal{L}} \widetilde{\mathsf{D}}^{\pm} 
    \right) \bigg\vert_{\text{Ising}}.
\end{align}
To summarize, we demonstrate that the anomalous algebra between translation and fermion parity in the enlarged Majorana system descends, after the GSO projection, to non-invertible symmetries $\widetilde{\mathsf{D}}^\pm$ of the bosonized open quantum system.

\subsection{Bosonization for Odd $L=2N+1$}\label{sec:bosonization_odd}
In this section, we bosonize the dissipative Majorana chain with odd system size $L=2N+1$. In Sec. \ref{sec:anomaly_RNS_odd_with_defect} we demonstrated that the Lindbladian Majorana chain, unlike the closed Majorana chain, exhibits an anomalous symmetry structure. One might therefore expect the corresponding bosonized description to exhibit some non-invertible symmetry. As we will see, however, no non-invertible symmetry arises in this case. The reasons are twofold. First, since the two strong fermion parity symmetries $(-1)^{F^\pm}$ anticommute with each other, we cannot simultaneously project onto definite eigenvalues of both. Second, requiring the ket- and bra-sector spin operators to commute with one another make the strong fermion parities to be eliminated from the sector Hamiltonians. Nevertheless, the bosonization, without imposing such a projection, naturally leads to an open-system Ising model with a Kramers--Wannier duality defect.

We begin with extending the Jordan--Wigner transformation used as in Eqs. (\ref{eq:JW_1})-(\ref{eq:JW_4}) by introducing an additional qubit at site $2N+1$,
\begin{align}
    &\chi_{2N+1}^{+}= - \left( \prod_{k=1}^{2N} Z_{k} \right) X_{2N+1}\\
    &\chi_{2N+1}^{-}= \left( \prod_{k=1}^{2N} Z_{k} \right) Z_{2N+1}.
\end{align}
Direct multiplication gives
\begin{align}
    &\chi_{2j-1}^+ \chi_{2j}^+ = -i Z_j\\
    &\chi_{2j-1}^- \chi_{2j}^- = i Z_{N+j}\\
    &\chi_{2N+1}^+ \chi_{2N+1}^- = i Y_{2N+1},
\end{align}
where $j=1,2,\cdots,N$, and
\begin{align}
    &\chi_{2j}^+ \chi_{2j+1}^+ = - i X_j X_{j+1},\\
    &\chi_{2j}^- \chi_{2j+1}^- = i X_{N+j} X_{N+j+1},
\end{align}
where $j=1,2,\cdots,N-1$. The terminal and boundary bilinears are
\begin{align}
    &\chi_{2N}^+ \chi_{2N+1}^+ = - i X_{N} X_{2N+1} \left(\prod_{k=N+1}^{2N} Z_k\right),\\
    &\chi_{2N+1}^+ \chi_1^+ = \left(\prod_{k=1}^{2N} Z_k\right) X_{2N+1} X_1,\\
    &\chi_{2N}^- \chi_{2N+1}^- = i X_{2N} Z_{2N+1},\\
    &\chi_{2N+1}^- \chi_1^- = \left(\prod_{k=N+1}^{2N} Z_k\right) Z_{2N+1} X_{N+1}.
\end{align}

The Jordan--Wigner transformation can be inverted to express the Pauli operators in terms of the Majorana operators:
\begin{align}
    &Z_j = i \chi_{2j-1}^+ \chi_{2j}^+,\\
    &Z_{N+j} = - i \chi_{2j-1}^- \chi_{2j}^-,\\
    &X_j = - \left(
    \prod_{k=1}^{j-1}\left( i \chi_{2k-1}^+ \chi_{2k}^+ \right)
    \right)
    \chi_{2j-1}^+,\\
    &Y_j = \left(
    \prod_{k=1}^{j-1}\left( i \chi_{2k-1}^+ \chi_{2k}^+ \right)
    \right) \chi_{2j}^+,\\
    &X_{N+j} = \left(
    \prod_{k=1}^{N}\left( i \chi_{2k-1}^+ \chi_{2k}^+ \right)
    \right)
    \left(
    \prod_{k=1}^{j-1}\left( - i \chi_{2k-1}^- \chi_{2k}^- \right)
    \right) \chi_{2j-1}^-,\\
    &Y_{N+j} = \left(
    \prod_{k=1}^{N}\left( i \chi_{2k-1}^+ \chi_{2k}^+ \right)
    \right)
    \left(
    \prod_{k=1}^{j-1}\left( - i \chi_{2k-1}^- \chi_{2k}^- \right)
    \right) \chi_{2j}^-,
\end{align}
where $j=1,2,\cdots,N$, and
\begin{align}
    &Z_{2N+1} = \left(
    \prod_{k=1}^{N}\left( i \chi_{2k-1}^+ \chi_{2k}^+ \right)
    \right)
    \left(
    \prod_{k=1}^{N}\left( -i \chi_{2k-1}^- \chi_{2k}^- \right)
    \right) \chi_{2N+1}^-,\\
    &X_{2N+1} = - \left(
    \prod_{k=1}^{N}\left( i \chi_{2k-1}^+ \chi_{2k}^+ \right)
    \right)
    \left(
    \prod_{k=1}^{N}\left( -i \chi_{2k-1}^- \chi_{2k}^- \right)
    \right) \chi_{2N+1}^+,\\
    &Y_{2N+1} = -i \chi_{2N+1}^+ \chi_{2N+1}^-.
\end{align}

The strong fermion parities in Eqs. (\ref{eq:strong_fermion_parity_plus_odd}) and (\ref{eq:strong_fermion_parity_minus_odd}) then take the form,
\begin{align}
    &(-1)^{F^{+}}=\left[\prod_{l=1}^{N}
        \left(
        -i\chi^{-}_{2l-1}\chi^{-}_{2l}
        \right)
    \right]
    \chi_{2N+1}^{-}
    = \left( \prod_{k=1}^N Z_k \right) Z_{2N+1},\label{eq:strong_fermion_parity_plus_in_terms_of_Pauli}\\
    &(-1)^{F^{-}}=\left[\prod_{l=1}^{N}
        \left(
        i\chi^{+}_{2l-1}\chi^{+}_{2l}
        \right)
    \right]
    \chi_{2N+1}^{+}
    = - \left( \prod_{k=N+1}^{2N} Z_k \right) X_{2N+1},\label{eq:strong_fermion_parity_minus_in_terms_of_Pauli}
\end{align}
which satisfy
\begin{align}
    &\left\{(-1)^{F^+},X_j\right\}=0,\quad
    \left\{(-1)^{F^+},Y_j\right\}=0,\quad
    \left[(-1)^{F^+},Z_j\right]=0,\quad\\
    &\left[(-1)^{F^+},X_{N+j}\right]=\left[(-1)^{F^+},Y_{N+j}\right]=\left[(-1)^{F^+},Z_{N+j}\right]=0,\\
    &\left[(-1)^{F^-},X_j\right]=\left[(-1)^{F^-},Y_j\right]=\left[(-1)^{F^-},Z_j\right]=0,\\
    &\left\{(-1)^{F^-},X_{N+j}\right\}=0,\quad
    \left\{(-1)^{F^-},Y_{N+j}\right\}=0,\quad
    \left[(-1)^{F^-},Z_{N+j}\right]=0,\quad
\end{align}

Now, we define the ket/bra Pauli operators as the following,
\begin{align}
    &Z_j^+ \equiv Z_j,\\
    &Z_j^- \equiv \mathcal{J} Z_j^+ \mathcal{J}^{-1} = Z_{N+j},\\
    &X_j^+ \equiv (-1)^{F^-} X_j,\label{eq:parity_dressing_1}\\
    &X_j^- \equiv \mathcal{J} X_j^+ \mathcal{J}^{-1} = - (-1)^{F^+}
    \left( \prod_{k=1}^N Z_k \right) X_{N+j},\\
    &Y_j^+ \equiv (-1)^{F^-} Y_j,\\
    &Y_j^- \equiv \mathcal{J} Y_j^+ \mathcal{J}^{-1} = (-1)^{F^+}
    \left( \prod_{k=1}^N Z_k \right) Y_{N+j},\label{eq:parity_dressing_4}
\end{align}
where $j=1,2,\cdots,N$. As before, the ket- and bra-sector operators are related by the modular conjugation. The ket-sector operators obey the conventional algebra, for example $X_j^+ Y_j^+ = i Z_j^+$, whereas the bra-sector operators obey the reversed algebra, such as $X_j^- Y_j^- = -i Z_j^-$. The parity dressing in Eqs. (\ref{eq:parity_dressing_1})-(\ref{eq:parity_dressing_4}) is chosen to ensure that every ket-sector Pauli operator commutes every bra-sector Pauli operator, as required for bosonic tensor structure. Moreover, by this construction, both strong fermion parities commute with all these dressed Pauli operators.

We proceed to rewrite the no-defect ket Hamiltonian in terms of newly defined ket/bra Pauli operators,
\begin{align}
    H_0^+ =& -\frac{i}{2} \sum_{l=1}^{L-1} \chi_{l}^{+} \chi_{l+1}^{+} - \frac{i}{2} \chi_{L}^{+} \chi_{1}^{+}\\
    =& -\frac{1}{2} \sum_{j=1}^{N} Z_j - \frac{1}{2} \sum_{j=1}^{N-1} X_j X_{j+1} - \frac{1}{2} X_N X_{2N+1} \left( \prod_{k=N+1}^{2N} Z_k \right) - \frac{i}{2} \left( \prod_{k=1}^{2N} Z_k \right) X_{2N+1} X_1\\
    =& -\frac{1}{2} \sum_{j=1}^{N} Z_j - \frac{1}{2} \sum_{j=1}^{N-1} X_j X_{j+1} + \frac{1}{2} X_N (-1)^{F^-} + \frac{i}{2} \left( \prod_{k=1}^{N} Z_k \right) (-1)^{F^-} X_1\\
    =&-\frac{1}{2} \sum_{j=1}^{N} Z_j^+ -\frac{1}{2} \sum_{j=1}^{N-1} X_j^+ X_{j+1}^+ + \frac{1}{2} X_N^+ + \frac{i}{2} \left( \prod_{k=1}^{N}Z_k^+ \right) X_1^+,\label{eq:H0_plus_with_Pauli_plus}
\end{align}
where we have applied Eq. (\ref{eq:strong_fermion_parity_minus_in_terms_of_Pauli}) in the second equality. Surprisingly, unlike the case of even system size $L=2N$, the strong fermion parities do no appear at the end---they are absorbed into the dressed Pauli operators. This is indirectly because we request the commutation between the ket and bra sectors.
Similarly, the ket-sector Hamiltonian with the defect takes the following form,
\begin{align}
    H_\mathsf{G}^+ = & -\frac{i}{2} \sum_{l=1}^{L-1} \chi_{l}^{+} \chi_{l+1}^{+} + \frac{i}{2} \chi_{L}^{+} \chi_{1}^{+}\\
    =& -\frac{1}{2} \sum_{j=1}^{N} Z_j^+ -\frac{1}{2} \sum_{j=1}^{N-1} X_j^+ X_{j+1}^+ +
    \frac{1}{2} X_N^+ - \frac{i}{2} \left( \prod_{k=1}^{N}Z_k^+ \right) X_1^+.\label{eq:HG_plus_with_Pauli_plus}
\end{align}
Applying modular conjugation to Eqs. (\ref{eq:H0_plus_with_Pauli_plus}) and (\ref{eq:HG_plus_with_Pauli_plus}) gives the corresponding bra-sector Hamiltonians,
\begin{align}
    &H_0^- = -\frac{1}{2} \sum_{j=1}^{N} Z_j^- -\frac{1}{2} \sum_{j=1}^{N-1} X_j^- X_{j+1}^-
    + \frac{1}{2} X_N^- - \frac{i}{2} \left( \prod_{k=1}^{N}Z_k^- \right) X_1^-,\\
    &H_\mathsf{G}^- = -\frac{1}{2} \sum_{j=1}^{N} Z_j^- -\frac{1}{2} \sum_{j=1}^{N-1} X_j^- X_{j+1}^-
    + \frac{1}{2} X_N^- + \frac{i}{2} \left( \prod_{k=1}^{N}Z_k^- \right) X_1^-.
\end{align}

We then focus on the enlarged Hamiltonian, Eq. (\ref{eq:tildeH_plus_direct_sum_four_subspaces}),
\begin{align}
    \widetilde{H}^+ =& H_{\mathsf{G}}^+ \oplus H_{\mathsf{G}}^+ \oplus H_0^+ \oplus H_0^+\\
    =&-\frac{1}{2} \sum_{j=1}^{N} \widetilde{Z}_j^+ -\frac{1}{2} \sum_{j=1}^{N-1} \widetilde{X}_j^+ \widetilde{X}_{j+1}^+ + \frac{1}{2} \widetilde{X}_N^+
    - \eta^+ \frac{i}{2} \left( \prod_{k=1}^{N}\widetilde{Z}_k^+ \right) \widetilde{X}_1^+,
\end{align}
where
\begin{align}
    &\widetilde{Z}_j^+ = Z_j^+ \oplus Z_j^+ \oplus Z_j^+ \oplus Z_j^+ = I_2 \otimes I_2 \otimes Z_j^+,\\
    &\widetilde{X}_j^+ = X_j^+ \oplus X_j^+ \oplus X_j^+ \oplus X_j^+ = I_2 \otimes I_2 \otimes X_j^+.
\end{align}
Similarly,
\begin{align}
    \widetilde{H}^- &= H_{\mathsf{G}}^- \oplus H_0^- \oplus H_{\mathsf{G}}^- \oplus H_0^-\\
    &= -\frac{1}{2} \sum_{j=1}^{N} \widetilde{Z}_j^- -\frac{1}{2} \sum_{j=1}^{N-1} \widetilde{X}_j^- \widetilde{X}_{j+1}^- + \frac{1}{2} \widetilde{X}_N^-
    + \eta^- \frac{i}{2} \left( \prod_{k=1}^{N}\widetilde{Z}_k^- \right) \widetilde{X}_1^-,
\end{align}
with
\begin{align}
    &\widetilde{Z}_j^- = Z_j^- \oplus Z_j^- \oplus Z_j^- \oplus Z_j^- = I_2 \otimes I_2 \otimes Z_j^-,\\
    &\widetilde{X}_j^- = X_j^- \oplus X_j^- \oplus X_j^- \oplus X_j^- = I_2 \otimes I_2 \otimes X_j^-.
\end{align}
Notice that the ket and bra Pauli operators in the enlarged space are connected by the modular conjugation $\widetilde{\mathcal{J}}$ in Eq. (\ref{eq:tilde_J}).

To cast the enlarged Hamiltonian into a more familiar form, we interpret $\eta^{\pm}$ as the $Z$ operators of two auxiliary qubits at site $N+1$, one associated with each sector,
\begin{align}
    &\widetilde{Z}_{N+1}^+ \equiv \widetilde{\eta}^+ = \sigma_z \otimes I_2 \otimes I_{\text{spin}},\\
    &\widetilde{Z}_{N+1}^- \equiv \widetilde{\eta}^- = I_2 \otimes \sigma_z \otimes I_{\text{spin}},
\end{align}
where $\sigma_{x,y,z}$ are conventional 2 by 2 Pauli matrices. We complete the ket-sector Pauli algebra by defining
\begin{align}
    &\widetilde{X}_{N+1}^+ = \sigma_x \otimes I_2 \otimes I_{\text{spin}},\\
    &\widetilde{Y}_{N+1}^+ = \sigma_y \otimes I_2 \otimes I_{\text{spin}}.
\end{align}
Applying the modular conjugation $\widetilde{\mathcal{J}}$ gives the bra correspondence,
\begin{align}
    &\widetilde{X}_{N+1}^- = I_2 \otimes \sigma_x \otimes I_{\text{spin}},\\
    &\widetilde{Y}_{N+1}^- = - I_2 \otimes \sigma_y \otimes I_{\text{spin}}.
\end{align}
One can easily check that the $N+1$ ket-sector Pauli operators satisfy the conventional algebra, the $N+1$ bra-sector Pauli operators follow the reversed algebra, and the two sets commute with each other. The Hamiltonians then become
\begin{align}
    \widetilde{H}^+ = &-\frac{1}{2} \sum_{j=1}^{N} \widetilde{Z}_j^+ -\frac{1}{2} \sum_{j=1}^{N-1} \widetilde{X}_j^+ \widetilde{X}_{j+1}^+ + \frac{1}{2} \widetilde{X}_N^+
    - \frac{i}{2} \widetilde{Z}_{N+1}^+ \left( \prod_{k=1}^{N}\widetilde{Z}_k^+ \right) \widetilde{X}_1^+,\label{eq:tilde_H_plus_with_tilde_Pauli_plus}\\
    \widetilde{H}^- =& -\frac{1}{2} \sum_{j=1}^{N} \widetilde{Z}_j^- -\frac{1}{2} \sum_{j=1}^{N-1} \widetilde{X}_j^- \widetilde{X}_{j+1}^- + \frac{1}{2} \widetilde{X}_N^-
    + \frac{i}{2} \widetilde{Z}_{N+1}^- \left( \prod_{k=1}^{N}\widetilde{Z}_k^- \right) \widetilde{X}_1^-.\label{eq:tilde_H_minus_with_tilde_Pauli_minus}
\end{align}
We next perform the following redefinition of the ket-sector spin variables,
\begin{align}
    &\widehat{X}_j^+ \equiv \widetilde{X}_j^+ \widetilde{X}_{N+1}^+,\\
    &\widehat{Y}_j^+ \equiv \widetilde{Y}_j^+ \widetilde{X}_{N+1}^+,\\
    &\widehat{Z}_j^+ \equiv \widetilde{Z}_j^+,\\
    &\widehat{X}_{N+1}^+ = -\widetilde{X}_{N+1}^+,\\
    &\widehat{Y}_{N+1}^+ = - \left( \prod_{k=1}^{N}\widetilde{Z}_k^+ \right) \widetilde{Y}_{N+1}^+,\\
    &\widehat{Z}_{N+1}^+ = \left( \prod_{k=1}^{N}\widetilde{Z}_k^+ \right) \widetilde{Z}_{N+1}^+,
\end{align}
where $j=1,2,\cdots,N$. These operators satisfy the conventional Pauli algebra for $N+1$ spins. The ket-space Hamiltonian in Eq. (\ref{eq:tilde_H_plus_with_tilde_Pauli_plus}) then takes the form,
\begin{align}
    \widetilde{H}^+ = - \frac{1}{2} \sum_{j=1}^N \widehat{Z}_j^+ - \frac{1}{2} \sum_{j=1}^N \widehat{X}_j^+ \widehat{X}_{j+1}^+ - \frac{1}{2} \widehat{X}_1^+ \widehat{Y}_{N+1}^+.\label{eq:tilde_H_plus_with_hatted_Pauli}
\end{align}
Using the modular conjugation, for instance $\tilde{\mathcal{J}} \widehat{Z}_j^+ \mathcal{J}^{-1} = \widehat{Z}_j^-$, the bra-sector Hamiltonian in Eq. (\ref{eq:tilde_H_minus_with_tilde_Pauli_minus}) takes the form,
\begin{align}
    \widetilde{H}^- = - \frac{1}{2} \sum_{j=1}^N \widehat{Z}_j^- - \frac{1}{2} \sum_{j=1}^N \widehat{X}_j^- \widehat{X}_{j+1}^- - \frac{1}{2} \widehat{X}_1^- \widehat{Y}_{N+1}^-.\label{eq:tilde_H_minus_with_hatted_Pauli}
\end{align}
Together, Eqs. (\ref{eq:tilde_H_plus_with_hatted_Pauli}) and (\ref{eq:tilde_H_minus_with_hatted_Pauli}) provide the ket and bra copies of a transverse-field Ising chain with a Kramers--Wannier duality defect on the bond $(N+1,1)$\cite{aasen2016topological}. For the collective energy dephasing jump in Eq. (\ref{eq:L0_collective_density_dephasing_jump_op}), $L_0 \propto H_0$, so the same construction also yields a corresponding bosonized Lindbladian with a Kramers--Wannier duality defect.

\section{Conclusion}\label{sec:conclusion}

In this paper, we studied symmetries and anomalies of dissipative Majorana chains by extending the lattice anomaly framework for Hamiltonian Majorana chains to Lindblad dynamics. The central idea is to use the operator-state mapping and regard the Lindbladian as a non-Hermitian operator acting on the doubled Hilbert space. This formulation allows us to analyze symmetries of open quantum systems in parallel with those of Hamiltonian systems, while also revealing structures that are intrinsic to open quantum systems, such as strong and weak symmetries and modular conjugation.

Our lattice analysis compared the symmetry algebras for even and odd chain lengths, with and without a weak fermion-parity defect. In the doubled Hilbert space, the ket and bra contours admit strong fermion parities, strong translations, and a weak fermion parity and weak translation acting diagonally on the two contours. We showed that these symmetry operators obey projective algebras that are direct open-system analogs of the Majorana-chain lattice anomaly. Characteristic results are that, for even length without a defect, untwisted sector exhibits projective relations involving both strong and weak symmetries,
\begin{align}
(-1)^{F^\pm}T_\text{RR}^\text{w}
=-T_\text{RR}^\text{w}(-1)^{F^\pm},
\qquad
(-1)^F_\text{w} T_\text{RR}^\pm
=-T_\text{RR}^\pm(-1)^{F_\text{w}}.\label{eq:conclusion_main_eqs_1}
\end{align}
This mixed algebra involves a weak symmetry and strong symmetries, and therefore has no direct counterpart in ordinary Hamiltonian systems. Comparing the untwisted and twisted sectors reveals changes in the symmetry algebras, providing a defect-based diagnosis of the LSM-type anomaly.

Odd-length chains reveal an additional feature of the doubled formulation: fermion-parity that cannot be implemented within a single irreducible closed-chain representation can be represented by operators in the doubled space. This allows weak fermion-parity defects to be introduced and their symmetry algebras to be examined. The odd-chain translation phases retain the modulo-$2$ lattice obstruction, while the defect sector exhibits further projective relations as in Eq. (\ref{eq:conclusion_main_eqs_1}).

We further matched these lattice symmetry algebras with the continuum doubled Majorana CFT in the RR, NSNS, NSR, and RNS sectors. The microscopic strong and weak translations emanate to continuum translations dressed by left-chiral fermion parities,
\begin{align}
    &T^\pm \sim (-1)^{F_\text{L}^\pm} e^{\pm i 2\pi P^\pm / L},\\
    &T^{\text{w}} \sim (-1)^{F_\text{L}^+} (-1)^{F_\text{L}^-} e^{i 2\pi P^+ / L} e^{-i 2\pi P^- / L},\label{eq:conclusion_main_eqs_2}
\end{align}
where $\sim$ denotes equality up to sector-dependent phases. Using these dictionaries, we derived the sector-dependent momentum quantization and (projective) relations between fermion parities. The anomalous minus signs in the RR sector and the fractional winding phases encoding modulo-$8$ classes in the NSR and RNS sectors persist in this description. Comparing RR with NSNS and NSR with RNS also reproduces the changes in projective commutation rules induced by the weak fermion-parity defect, providing a continuum realization of the anomalous strong and weak symmetry structure.

Finally, we developed a lattice bosonization of the dissipative Majorana chain. For even length, we combined the defect and no-defect sectors and imposed GSO projections on both the ket and bra contours. Constructing spin operators that preserve the projected subspace yields an Ising Lindbladian (for the parity-even jump). The Majorana translations themselves do not preserve this subspace, and this issue can be resolved by retaining only the compatible components through a projection, giving
\begin{align}
    \widetilde{\mathsf{D}}^\pm &= \frac{1+\widetilde{\eta}^\pm}{2} \widetilde{T}^\pm.
\end{align}
After restricting to the Ising subspace, the resulting symmetries $\mathsf{D}^\pm$ have nontrivial kernels and implement Kramers--Wannier duality separately on the ket and bra contours, exchanging the Ising field and bond terms. Thus, these non-invertible operators make
\begin{align}
    \mathsf{D}^\pm \mathcal{L} = \mathcal{L} \mathsf{D}^\pm.
\end{align}

For odd length, we constructed commuting ket- and bra-sector spin operators by parity dressing, which absorbs the strong parities and removes their explicit appearance in the sector Hamiltonians and jump operators. Performing a further spin redefinition brings the enlarged Hamiltonians and jump terms to a ket and bra copies of a transverse-field Ising chain with Kramers--Wannier duality defect. Unlike the even case, the anomalous relations do not produce a non-invertible symmetry.

There are several directions for future work. First, it would be important to complete the bosonic formulation of the open-chain anomaly and derive the full non-invertible symmetry algebra of the open Ising Lindbladian. Second, the relation between the projective symmetry algebra and constraints on the Liouvillian spectrum, steady-state degeneracy, and dissipative gap should be clarified more systematically, for example through twisted dissipative form factors. Third, the continuum analysis should be extended beyond the free Majorana CFT to interacting fermionic conformal field theories, such as the superconformal tricritical Ising model\cite{PhysRevLett.115.166401}. We expect that the interplay among lattice translation, fermion parity, modular conjugation, and non-invertible symmetry provides a useful framework for understanding anomalies and constraints in open quantum many-body systems.

\section*{Acknowledgements}
We have benefited from discussions with Shinsei Ryu, Kohei Kawabata , Ryusuke Hamazaki while preparing this manuscript.
H.-H.T. was supported by JST PRESTO (Grant No.~JPMJPR2359) and JSPS KAKENHI (Grant No.~24H00829).
T.N. was supported by MEXT KAKENHI (Grant No.~23K13094 and 24H00944) and by JST PRESTO (Grant No.~JPMJPR2359).

\clearpage
\bibliographystyle{utphys} 
\bibliography{manuscript}

\end{document}